\documentclass[11pt]{article}
\usepackage{jheppub}

\usepackage{amsfonts,amssymb,amsmath,amsthm,url,graphicx,mathtools}
\usepackage{mathrsfs}
\usepackage{tikz}
\usepackage{url}
\usepackage{psfrag}

\numberwithin{equation}{section}

\DeclareMathOperator{\SO}{SO}

\newcommand{\Winf}[1]{\mathcal{W}_{\infty}[#1]}
\newcommand{\Winfcl}[1]{\mathcal{W}^{cl}_{\infty}[#1]}

\newcommand{\wg}[1]{\mathfrak{shs}^\sigma[#1]}

\newcommand{\be}{\begin{equation}}
\newcommand{\ee}{\end{equation}}

\newcommand{\W}{\mathcal{W}}

\DeclareMathOperator{\Cas}{Cas}

\DeclareMathOperator{\ch}{ch}

\numberwithin{equation}{section}

\def\be{\begin{equation}}
\def\ee{\end{equation}}

\title{The $\mathcal{N}=1$  algebra $\Winf{\mu}$ and its truncations}

\author{Constantin Candu and
Carl Vollenweider}

\affiliation{
Institut f\"ur Theoretische Physik, ETH Z\"urich \\
CH-8093 Z\"urich, Switzerland}

\emailAdd{canduc@itp.phys.ethz.ch}
\emailAdd{carlv@itp.phys.ethz.ch}

\abstract{
The main objective of this work is to 
construct and classify the most general classical and quantum $\mathcal{N}=1$ $\mathcal{W}_\infty$-algebras
generated by the same spins as the singlet algebra of $N$ fermions and $N$ bosons in the vector representation of $\mathrm{O}(N)$ in the $N\to\infty$ limit.
This type of algebras appears in a recent $\mathcal{N}=1$ version of the minimal model holography.
Our analysis strongly suggests that there is a one parameter family $\mathcal{W}_\infty[\mu]$ of such algebras at every given central charge.
Relying on this assumption, we  identify  various truncations of $\mathcal{W}_\infty[\mu]$ with, on the one hand,  (orbifolds of) the Drinfel'd-Sokolov reductions of the Lie superalgebras
$B(n,n)$, $B(n-1,n)$, $D(n,n)$ and $D(n+1,n)$, and, on the other hand, (orbifolds of) three $\mathcal{N}=1$ cosets.
After a closer inspection we show that these cosets can be realized as a Drinfel'd-Sokolov reduction of $B(n,n)$,  $D(n,n)$ and $D(n+1,n)$.
We then discuss the implications of our findings for the quantum version of the $\mathcal{N}=1$ minimal model holography.
}

\date{\today}

\allowdisplaybreaks

\begin{document}
\maketitle


\section{Introduction}

The importance of infinitely generated $\mathcal{W}$-algebras has recently reemerged in the context of the   minimal model holography~\cite{Gaberdiel:2010pz}.
At present there are four versions of  this type of AdS$_3$/CFT$_2$ dualities, claiming a holographic realization of the unitary coset models \cite{Gaberdiel:2010pz, Gaberdiel:2011nt, Ahn:2011pv, Creutzig:2011fe, Creutzig:2012ar}  
\begin{equation}
\frac{\mathfrak{su}(N)_k\oplus\mathfrak{su}(N)_1}{\mathfrak{su}(N)_{k+1}}\ ,\quad \frac{\mathfrak{so}(N)_k\oplus\mathfrak{so}(N)_1}{\mathfrak{so}(N)_{k+1}}\ ,\quad \frac{\mathfrak{su}(N+1)_k\oplus\mathfrak{so}(2N)_1}{\mathfrak{su}(N)_{k+1}\oplus \mathfrak{u}(1)}\ ,\quad \frac{\mathfrak{so}(N+1)_k\oplus\mathfrak{so}(N)_1}{\mathfrak{so}(N)_{k+1}}\ ,
\label{eq:4cosets}
\end{equation}
in terms of various  AdS$_3$ higher spin (super)gravity theories found by Prokushkin and Vasiliev \cite{Prokushkin:1998bq}.
The first two dualities are non-supersymmetric, while the last two are $\mathcal{N}=2$ and $\mathcal{N}=1$ supersymmetric, respectively.
Contact with classical (super)gravity was initially made in the large $N$ 't~Hooft like limit
\begin{equation}
N,k \to \infty\quad   \text{with}\quad \lambda = \frac{N}{N+k} \quad \text{held fixed} \ ,
\label{eq:thooft_gen}
\end{equation}
in which the coset central charge diverges as $c\propto N$.
This behavior is characteristic of vector like models and, indeed, in many respects the above dualities
can be viewed as lower-dimensional analogs of the
original Klebanov-Polyakov conjecture~\cite{Klebanov:2002ja}, which states that the singlet sector of the $\mathrm{O}(N)$ vector model in 3d is dual in the large $N$ limit to an AdS$_4$ higher spin gravity theory of Vasiliev \cite{Vasiliev:2003ev}.
For recent reviews of vector model/higher spin dualities see \cite{Gaberdiel:2012uj, Giombi:2012ms, Ammon:2012wc, Jevicki:2012fh, Jin:2013lqa}.

The Vasiliev theories  dual to \eqref{eq:4cosets} are classical theories of AdS$_3$ (super)gravity coupled to massive matter and an infinite tower of higher spin gauge fields.
Their gauge sector can be reformulated as a  Chern-Simons (CS) theory \cite{Blencowe:1988gj} in the same way as usual AdS$_3$ (super)gravity with $\mathcal{N}$ supersymmetries  can be reformulated as (two copies of) an $\mathfrak{osp}(\mathcal{N}|2)$ CS theory~\cite{Achucarro:1987vz, Witten:1988hc}.
In this reformulation the gauge fields  take values in (two copies of) certain 
infinite dimensional, infinite rank, one parameter families of so-called higher spin Lie (super)algebras~\cite{feigin, post, Pope:1989sr, Bordemann:1989zi, Fradkin:1990qk}
\begin{equation}
\mathfrak{sl}(2)\subset\mathfrak{hs}[\mu]\ , \quad \mathfrak{sl}(2)\subset\mathfrak{hs}^{(e)}[\mu]\ , \quad \mathfrak{osp}(2|2)\subset\mathfrak{shs}[\mu]\ , \quad \mathfrak{osp}(1|2)\subset\mathfrak{shs}^\sigma[\mu]\ ,
\label{eq:list_hs}
\end{equation}
which  come equipped with an  $\mathfrak{osp}(\mathcal{N}|2)$ embedding singling out the (super)graviton multiplet and defining  the asymptotic AdS$_3$ geometry.
The topological nature of the  graviton and higher spin fields, made manifest by the CS reformulation, implies that their dynamics is localized at the asymptotic boundary of AdS$_3$.
This dynamics is governed by the algebra of asymptotic symmetries \emph{\`a la} Brown \& Henneaux \cite{Brown:1986nw}, i.e.\ gauge symmetries respecting the  asymptotic AdS$_3$ geometry.
At a technical level, the asymptotic symmetries can be identified with a classical Poisson bracket $\mathcal{W}_\infty$-algebra  \cite{Henneaux:2010xg, Campoleoni:2010zq, Campoleoni:2011hg, Gaberdiel:2011wb, Henneaux:2012ny, Hanaki:2012yf} 
constructed as the  classical Drinfel'd-Sokolov (DS) reduction  of the respective higher spin algebra w.r.t.\ the  $\mathfrak{osp}(\mathcal{N}|2)$ (super)gravity embedding~\cite{Balog:1990mu, Delduc:1991sg}.
A good control of it gives direct access to all  graviton and higher spin correlation functions.
At a conceptual level, the asymptotic symmetries establish a natural bridge between the bulk (super)gravity theory and its dual  CFT.
The relationship can be understood as follows.

Finitely generated quantum $\mathcal{W}$-algebras have their origin in  CFTs, where  they appear in the form of chiral algebras, i.e.\ extensions of the (super)Virasoro algebra by local holomorphic currents \cite{Bouwknegt:1992wg}.
Taking certain limits in which the central charge diverges one can obtain  $\mathcal{W}_\infty$-algebras.
In this way, extensive evidence was presented in  \cite{Gaberdiel:2012, Candu:2012tr, Candu:2012ne} that the chiral algebras of the  first three 
cosets in~\eqref{eq:4cosets} become in the 't~Hooft  limit~\eqref{eq:thooft_gen}
classical  $\mathcal{W}_\infty$-algebras isomorphic to the algebras of asymptotic symmetries of their conjectured bulk duals if  the parameter $\mu$ of the higher spin algebras~\eqref{eq:list_hs} is identified with $\lambda$.
An equally important piece of evidence, valid under certain assumptions about the emergence of null states in the 't Hooft limit, but this time available for all four cosets in~\eqref{eq:4cosets}, is the matching between the 1-loop partition function of the bulk theory
and the properly regularized 't~Hooft limit of the CFT partition function~\cite{Gaberdiel:2011zw, Gaberdiel:2011nt, Candu:2012jq, Creutzig:2012ar}.
More evidence, mostly available in the first case,
is provided by the  matching of correlation functions of dual pairs~\cite{Chang:2011mz, Papadodimas:2011pf, Ahn:2011by, Ammon:2011ua, Chang:2011vka, Creutzig:2012xb, Moradi:2012xd, Hijano:2013fja}, the CFT interpretation of higher spin black holes~\cite{Kraus:2011ds, Gaberdiel:2012yb, Datta:2013qja} and the construction of classical geometries
 corresponding to CFT primaries~\cite{Castro:2011iw, Perlmutter:2012ds, Datta:2012km,  Hikida:2012eu, Chen:2013oxa}.

These checks of the holography are strictly speaking valid only in the classical regime.
At present, the only known way to check the holographic duality in the quantum gravity regime consists of comparing the chiral algebra of the CFT at finite $N$ and $k$ with the quantization of the classical $\mathcal{W}_\infty$-algebra of asymptotic symmetries of the bulk theory.
The procedure to carry out this quantization was first explained in \cite{Gaberdiel:2012} on the example of the first coset
in \eqref{eq:4cosets} and then successfully applied to the next two cosets \cite{Candu:2012tr, Candu:2012ne}.
Every one of these analysis strongly suggested that the quantization procedure is unique, up to a discrete ambiguity. With this basic assumption one can then show the existence of an isomorphism between the chiral algebras of the CFT~\eqref{eq:4cosets} at finite $N$ and $k$ and a truncation of the  quantum $\mathcal{W}_\infty$-algebra of asymptotic symmetries arising at a precise value of $\mu$ as a function of $N$ and $k$.\footnote{In fact, the proof of the agreement of symmetries in the 't~Hooft limit passes through the quantum analysis.}
One can interpret this fact as the first convincing sign that holography continues to hold in the quantum regime.

In this paper we shall consider the last duality in \eqref{eq:4cosets}, which is the most recent and least studied one.
In particular, there is no (direct) evidence that the symmetry algebras on both sides agree, even in the 't~Hooft limit.
The primary goal of this paper is to quantize the algebra of asymptotic symmetries of the higher spin supergravity  theory and check the agreement with the symmetries of the coset at finite $N$ and $k$ and in the 't~Hooft limit, all of this following the same strategy as in \cite{Gaberdiel:2012, Candu:2012tr, Candu:2012ne}.
According to~\cite{Creutzig:2012ar}, the algebra of asymptotic symmetries is an $\mathcal{N}=1$ $\mathcal{W}_\infty$-algebra
generated by  $\mathcal{N}=1$ multiplets with Virasoro  spins
\begin{align}\label{eq:Winf_spec}
(\tfrac{3}{2}, 2)\ ,\ (2,\tfrac{5}{2})\ ,\ (\tfrac{7}{2},4)\ ,\ (4,\tfrac{9}{2})\ ,\ (\tfrac{11}{2},6)\ , \ (6,\tfrac{13}{2})\ ,\ (\tfrac{15}{2},8)\ ,
\ (8,\tfrac{17}{2})\ ,\ \dots
\end{align}
where the first supermultiplet corresponds to the $\mathcal{N}=1$ Virasoro algebra itself. Every $\mathcal{N}=1$ multiplet appears only once.
As we shall see, the singlet algebra of $N$ fermions and $N$ bosons in the vector representation of $\mathrm{O}(N)$ is generated in the limit $N\to\infty$ by the same  list of spins~\eqref{eq:Winf_spec}.
Therefore, $\mathcal{N}=1$ $\mathcal{W}_\infty$-algebras of this type naturally appear in $\mathcal{N}=1$ vector type holographic dualities.
From now on,  $\mathcal{W}_\infty$ will always denote an $\mathcal{N}=1$ $\mathcal{W}$-algebra generated by the collection of spins~\eqref{eq:Winf_spec}.

In sec.~\ref{sec:winf} we shall present extensive evidence (based on the associativity of low level OPEs) that the space of the most general quantum and classical $\mathcal{W}_\infty$-algebras is parametrized by only one parameter in addition to the central charge. This is our first basic assumption that we adopt throughout the paper.
Later this additional parameter will be  identified with the parameter $\mu$ of the higher spin algebra $\mathfrak{shs}^\sigma[\mu]$ of the bulk theory.
In order to compare the quantum $\mathcal{W}_\infty$-algebra with the chiral algebra of the coset we need to study its truncations.
This study  is based on the properties of the most degenerate representations of $\mathcal{W}_\infty$, called minimal representations (and defined by a Virasoro character).
In sec.~\ref{sec:min} we present extensive evidence (based on the associativity of low level OPEs) that the minimal representations exist and are uniquely determined by their conformal dimension.
This will be our second basic assumption, that we also adopt throughout the paper.
In sec.~\ref{sec:tr} we identify the DS reductions of the Lie superalgebras  $B(n,n)$ and $B(n-1,n)$, and $\mathbb{Z}_2$ orbifolds of the  DS reductions of  $D(n,n)$ and $D(n+1,n)$ with truncations of $\mathcal{W}_\infty$, both in the classical and quantum cases.
In sec.~\ref{sec:cosets}, we consider the two cosets\footnote{Our convention for the level of $\mathfrak{sp}(2n)_{k}$ is twice the usual one, hence $\mathfrak{sp}(2n)_{-1}$ stands for $n$ $\beta\gamma$-systems.}
\begin{equation}
\frac{\mathfrak{so}(N+1)_k\oplus\mathfrak{so}(N)_1}{\mathfrak{so}(N)_{k+1}}\ ,\qquad \frac{\mathfrak{osp}(1|2n)_k\oplus\mathfrak{sp}(2n)_{-1}}{\mathfrak{sp}(2n)_{k-1}}\ .
\label{eq:n1_cosets}
\end{equation}
First, we prove that their chiral algebras (in the first case only a $\mathbb{Z}_2$ orbifold) are truncations of $\mathcal{W}_\infty$, thus achieving our goal of matching
the symmetries of the bulk and boundary theories.
A closer inspection then reveals that the chiral algebra  of the first coset can be identified with the 
DS reduction of $D(n,n)$ or $D(n+1,n)$, depending on the parity of $N$, while that of the second coset can be identified with the DS reduction of $B(n,n)$.

There are  also five appendices.
We shall only mention app.~\ref{shssigma}, where we define the higher spin algebra $\wg{\mu}$, prove its truncation properties and discuss its simplest representations, and 
app.~\ref{app:one_more}, where we write down the first few Poisson brackets of the DS reduction of $\wg{\mu}$.




\section{\texorpdfstring{$\mathcal{N}=1$ $\mathcal{W}_\infty$}{N=1 Winfty}-algebras}
\label{sec:winf}


In the following we shall consider the most general quantum $\mathcal{W}_\infty$-algebra 
with the same spectrum as \eqref{eq:Winf_spec}.
Notice that all $\mathbb{N}+\tfrac{1}{2}$ valued spins appear once, there are no odd spins and all even spins appear twice.
%

We shall present the algebra structure in terms of the OPEs of its generators, which we choose to be $\mathcal{N}=1$ primary multiplets.
The defining OPEs of the $\mathcal{N}=1$ Virasoro algebra, generated by the Virasoro tensor $T(z)$ and the
supercurrent $G(z)$, are recalled in app.~\ref{sec:n=1}.
We shall denote the $\mathcal{N}=1$ multiplet $(s,s+\tfrac{1}{2})$ appearing in \eqref{eq:Winf_spec} by $W^{(s)}$ and reserve the notation $W^{s\,0}$ for its $\mathcal{N}=1$ primary
Virasoro component and $W^{s\,1}=G_{-\frac{1}{2}}W^{s\,0}$ for its primary superpartner.
The defining OPEs of $\mathcal{N}=1$ primary multiplets are recalled in app.~\ref{sec:n=1}.

Our analysis follows closely the approach of \cite{Gaberdiel:2012, Candu:2012tr, Candu:2012ne}, where other types of infinite $\mathcal{W}$-algebras were considered.
This approach is an iterative algorithm consisting of the following steps: (i) make  the most general ansatz for the OPEs $W^{(s_1)}\times W^{(s_2)}$ with $s_1+s_2$ lower or equal to a certain value called level, (ii)
 impose and solve  the associativity constraints of these OPEs and (iii) increase the level and repeat the whole procedure  up to the desired level.
From the analysis of the associativity constraints, we observe that, at least up to the level considered, all structure constants of the $\mathcal{W}_\infty$-algebra can be uniquely and non-redundantly expressed in terms of the central charge $c$ and one additional continuous parameter. Hence, we conjecture that the space of most general $\mathcal{W}_\infty$-algebras is parametrized by a single continuous parameter in addition to $c$. Finally, we find an interpretation of this parameter in terms of a
classical Drinfel'd-Sokolov reduction by studying the classical  limit of the algebra.


\subsection{Construction}\label{sec:constr}


It is a well-known fact (see~\cite{Belavin:1984vu, Blumenhagen:1991nm}) that superconformal symmetry fixes the coefficients 
of all fields 
appearing in the OPEs $W^{s_1\,\alpha_1}(z_1)W^{s_2\,\alpha_2}(z_2)$ 
in terms of the coefficients of $\mathcal{N}=1$ primaries only. Notice that imposing superconformal symmetry is equivalent to requiring
associativity of the OPEs $T(z)W^{s_1\,\alpha_1}(z_1)W^{s_2\,\alpha_2}(z_2)$ and $G(z)W^{s_1\,\alpha_1}(z_1)W^{s_2\,\alpha_2}(z_2)$.
Moreover, the singular part of the OPEs fully determines the regular part by  the Jacobi identities~\cite{Thielemans:1994, Thielemans:1991}.
For these two reasons, an ansatz for the OPEs $W^{s_1\,\alpha_1}(z_1)W^{s_2\,\alpha_2}(z_2)$, which we compactly denote by $W^{(s_1)}\times W^{(s_2)}$,
is already unambiguously specified  after the contribution of $\mathcal{N}=1$ primaries to the singular part is written down.

Using these conventions, we would now like to write down the most general ansatz for the OPE $W^{(s_1)}\times W^{(s_2)}$.
This ansatz assumes that every $\mathcal{N}=1$ primary (simple or composite) multiplet of the algebra
that has a spin small enough to appear in the singular part will  actually appear. Thus, we need to know how many 
$\mathcal{N}=1$ primary fields there are in the $\mathcal{W}_\infty$-algebra at every given spin. These can be computed by decomposing the vacuum character of the $\W_\infty$-algebra
\begin{equation}\label{eq:ch_vac}
\chi_\infty (q) = \mathrm{Tr}_{\W_\infty}q^{L_0}=
\frac{\prod_{s\in\mathbb{N}+\frac{1}{2}}\prod_{n=0}^{\infty}(1+q^{n+s})}{\prod_{s\in2\mathbb{N}}\prod_{n=0}^\infty(1-q^{n+s})^2}=\chi_0(q)+\sum_{h\in\frac{1}{2}\mathbb{N}}d(h)
\chi_h(q)
\ ,
\end{equation}
in terms of the $\mathcal{N}=1$ Virasoro vacuum character and characters of $\mathcal{N}=1$ highest weight representations of conformal dimension $h$
\begin{equation}
\chi_{0} (q) = \prod_{n=1}^\infty
\frac{1+q^{n+\frac{1}{2}}}{1-q^{n+1}}  \ ,\qquad \chi_{h}(q) = q^h \, \prod_{n=0}^\infty
\frac{1+q^{n+\frac{1}{2}}}{1-q^{n+1}} \ .\label{eq:n1_chs}
\end{equation}
The integers $d(h)$ in eq.~\eqref{eq:ch_vac} are the number of $\mathcal{N}=1$ primaries at conformal dimension $h$ that we were looking for.
Expanding the generating function
\be
D(q) = \sum_{h\in\frac{1}{2}\mathbb{N}}
d(h) \, q^h = 
\frac{1-q}{1+q^{\frac{1}{2}}} \left( \frac{\chi_\infty(q)}{\chi_0(q)} -1 \right) \ 
\ee
we get
\be\label{eq:p2_series}
D(q) = q^2 + q^{\frac{7}{2}} + 2 q^4 + 3 q^{\frac{11}{2}} + 5 q^6 + 2 q^{\frac{13}{2}} + 2 q^7 + \cdots \ .
\ee

The most general ansatz for the first few non-trivial OPEs of $\W_\infty$ is
\begin{align} \label{ope22}
W^{(2)} \times W^{(2)} &\sim n_2 I + c_{22}^2 W^{(2)} + c_{22}^{\frac{7}{2}} W^{(\frac{7}{2})} \ , \\ 
W^{(2)} \times W^{(\frac{7}{2})} &\sim c_{2 \frac{7}{2}}^2 W^{(2)} + c_{2 \frac{7}{2}}^{\frac{7}{2}} W^{(\frac{7}{2})}  \label{ope27h}
+ c_{2 \frac{7}{2}}^{4} W^{(4)} + a_{2 \frac{7}{2}}^{4} A^{(4)} \ ,\\ \notag
W^{(2)} \times W^{(4)} &\sim c_{2 4}^2 W^{(2)} + c_{2 4}^{\frac{7}{2}} W^{(\frac{7}{2})} 
+ c_{2 4}^{4} W^{(4)} + a_{2 4}^{4} A^{(4)} + c_{2 4}^{\frac{11}{2}} W^{(\frac{11}{2})} \\ \label{ope24}
&\quad +a_{2 4}^{\frac{11}{2} ,1} A^{(\frac{11}{2},1)} + a_{2 4}^{\frac{11}{2},2} A^{(\frac{11}{2},2)} \ , \\ \notag
W^{(\frac{7}{2})} \times W^{(\frac{7}{2})} &\sim n_{\frac{7}{2}} I + c_{\frac{7}{2} \frac{7}{2}}^2 W^{(2)} + c_{\frac{7}{2} \frac{7}{2}}^{\frac{7}{2}} W^{(\frac{7}{2})} 
+ c_{\frac{7}{2} \frac{7}{2}}^{4} W^{(4)} + a_{\frac{7}{2} \frac{7}{2}}^{4} A^{(4)} \\ \notag
&\quad + c_{\frac{7}{2} \frac{7}{2}}^{\frac{11}{2}} W^{(\frac{11}{2})}  +a_{\frac{7}{2} \frac{7}{2}}^{\frac{11}{2} ,1} A^{(\frac{11}{2},1)} 
+ a_{\frac{7}{2} \frac{7}{2}}^{\frac{11}{2},2} A^{(\frac{11}{2},2)} 
+ c_{\frac{7}{2} \frac{7}{2}}^{6} W^{(6)} \\ \notag
&\quad + a_{\frac{7}{2} \frac{7}{2}}^{6,1} A^{(6,1)} + a_{\frac{7}{2} \frac{7}{2}}^{6,2} A^{(6,2)}
+ a_{\frac{7}{2} \frac{7}{2}}^{6,3} A^{(6,3)} + a_{\frac{7}{2} \frac{7}{2}}^{6,4} A^{(6,4)} \\ \label{ope7h7h}
&\quad + a_{\frac{7}{2} \frac{7}{2}}^{\frac{13}{2},1} A^{(\frac{13}{2},1)} + a_{\frac{7}{2} \frac{7}{2}}^{\frac{13}{2},2} A^{(\frac{13}{2},2)} \ , \\ \notag
W^{(2)} \times W^{(\frac{11}{2})} &\sim c_{2 \frac{11}{2}}^2 W^{(2)} + c_{2 \frac{11}{2}}^{\frac{7}{2}} W^{(\frac{7}{2})} 
+ c_{2 \frac{11}{2}}^{4} W^{(4)} + a_{2 \frac{11}{2}}^{4} A^{(4)} \\ \notag
&\quad + c_{2 \frac{11}{2}}^{\frac{11}{2}} W^{(\frac{11}{2})}+a_{2 \frac{11}{2}}^{\frac{11}{2} ,1} A^{(\frac{11}{2},1)} 
+ a_{2 \frac{11}{2}}^{\frac{11}{2},2} A^{(\frac{11}{2},2)} 
+ c_{2 \frac{11}{2}}^{6} W^{(6)} \\ \notag
&\quad + a_{2 \frac{11}{2}}^{6,1} A^{(6,1)} + a_{2 \frac{11}{2}}^{6,2} A^{(6,2)}
+ a_{2 \frac{11}{2}}^{6,3} A^{(6,3)} + a_{2 \frac{11}{2}}^{6,4} A^{(6,4)} \\
&\quad + a_{2 \frac{11}{2}}^{\frac{13}{2},1} A^{(\frac{13}{2},1)} + a_{2 \frac{11}{2}}^{\frac{13}{2},2} A^{(\frac{13}{2},2)}\label{ope211h}
+ a_{2 \frac{11}{2}}^{7,1} A^{(7,1)} + a_{2 \frac{11}{2}}^{7,2} A^{(7,2)}  \ , \\ \notag
W^{(\frac{7}{2})} \times W^{(4)} &\sim c_{\frac{7}{2} 4}^2 W^{(2)} + c_{\frac{7}{2} 4}^{\frac{7}{2}} W^{(\frac{7}{2})} 
+ c_{\frac{7}{2} 4}^{4} W^{(4)} + a_{\frac{7}{2} 4}^{4} A^{(4)} \\ \notag
&\quad + c_{\frac{7}{2} 4}^{\frac{11}{2}} W^{(\frac{11}{2})}  +a_{\frac{7}{2} 4}^{\frac{11}{2} ,1} A^{(\frac{11}{2},1)} 
+ a_{\frac{7}{2} 4}^{\frac{11}{2},2} A^{(\frac{11}{2},2)} 
+ c_{\frac{7}{2} 4}^{6} W^{(6)} \\ \notag
&\quad + a_{\frac{7}{2} 4}^{6,1} A^{(6,1)} + a_{\frac{7}{2} 4}^{6,2} A^{(6,2)}
+ a_{\frac{7}{2} 4}^{6,3} A^{(6,3)} + a_{\frac{7}{2} 4}^{6,4} A^{(6,4)} \\ 
&\quad + a_{\frac{7}{2} 4}^{\frac{13}{2},1} A^{(\frac{13}{2},1)} + a_{\frac{7}{2} 4}^{\frac{13}{2},2} A^{(\frac{13}{2},2)}
+ a_{\frac{7}{2} 4}^{7,1} A^{(7,1)} + a_{\frac{7}{2} 4}^{7,2} A^{(7,2)} \ , \label{ope7h4}
\end{align}
where on the r.h.s.\ only the contribution of superprimary multiplets is displayed explicitly.
Here  $A^{(s,i)}$ denotes the $i$-th  composite superprimary multiplet at spin $s$.
The number of these composite fields is predicted by eq.~\eqref{eq:p2_series}.
Details on how to compute them and their rough form are presented in app.~\ref{sec:composite}.

The precise value of the maximal spin of a supermultiplet on the r.h.s.\ of eqs.~(\ref{ope22}--\ref{ope7h4}) is determined by requiring that it leads to a pole structure which is compatible  with $\mathcal{N}=1$ supersymmetry.
This means that the components of the supermultiplet must have a spin small enough so that both of them can contribute to at least some of the singular terms in the component OPEs $W^{s_1\, \alpha_1}(z_1) W^{s_2\, \alpha_2}(z_2)$.

Let us now explain our conventions for the structure constants in eqs.~(\ref{ope22}--\ref{ope7h4}).
If we denote by $c_{s_1s_2}^{s_3}(\alpha_1,\alpha_2)$ the coefficients of $W^{s_3\alpha_3}(w)$ in the singular part of the OPEs $W^{s_1\alpha_1}(z)W^{s_2\alpha_2}(w)$, then $\mathcal{N}=1$ supersymmetry relates these coefficients 
as follows
\begin{description}
\item[$i)$] $s_1+s_2-s_3 \in \mathbb{N}$
\begin{equation*}
c_{s_1s_2}^{s_3} (0,0) =\frac{  (-1)^{2 s_1 + 1}2 s_3 }{s_1 - s_2 -s_3} \, c_{s_1s_2}^{s_3} (0,1)
= \frac{2 s_3}{s_1 - s_2 +s_3} \, c_{s_1s_2}^{s_3} (1,0) =   \frac{(-1)^{2 s_1 + 1}}{s_1 + s_2 -s_3} \, c_{s_1s_2}^{s_3} (1,1) \ ,
\end{equation*}
\item[$ii)$] $s_1+s_2-s_3 \in \mathbb{N}-\tfrac{1}{2}$
\begin{equation}\label{eq:rel1}
c_{s_1s_2}^{s_3} (0,0) =  \frac{(-1)^{2 s_1 + 1}}{2 s_3} \, c_{s_1s_2}^{s_3} (0,1)
= \frac{1}{2 s_3} \, c_{s_1s_2}^{s_3} (1,0) = \frac{(-1)^{2 s_1 + 1}}{s_1 + s_2 +s_3 - \frac{1}{2}} \, c_{s_1s_2}^{s_3} (1,1) \ ,
\end{equation}
\end{description}
see \cite{Blumenhagen:1991nm}.
The structure constants $c_{s_1s_2}^{s_3}$ are defined as
\begin{equation}
c_{s_1s_2}^{s_3}=\begin{cases}
\;c_{s_1s_2}^{s_3}(0,0)& \text{ if }\;  s_1+s_2-s_3\in\mathbb{N}\; \text{ or }\; s_1+s_2-(s_3+\tfrac{1}{2})\in\mathbb{N}\ ,\\
\;c_{s_1s_2}^{s_3}(0,1)& \text{ if }\;  s_1+s_2-(s_3+\frac{1}{2})=0\ .
\end{cases}
\label{eq:def_c}
\end{equation}
%
%
The same conventions apply to the structure constants $a_{s_1s_2}^{s_3}$. Similarly, $n_s$ determines the normalization of the multiplet $W^{(s)}$
\begin{equation}\label{eq:ns}
W^{s\,0}(z) W^{s\,0}(w)= \frac{n_s}{(z-w)^{2s}}+\cdots \ \ ,\qquad W^{s\,1}(z) W^{s\,1}(w)= \frac{(-1)^{2s+1}2sn_s}{(z-w)^{2s+1}}+\cdots\ .
\end{equation}
%


\subsection{Classification}\label{sec:classif}


The structure constants in the ansatz~(\ref{ope22}--\ref{ope7h4}) are strongly constrained by the  associativity of the double OPEs $W^{s_1\, \alpha_1}(z_1)
W^{s_2\, \alpha_2}(z_2) W^{s_3\, \alpha_3}(z_3)$.
We refer the reader unfamiliar with associativity constraints to sec.~2.3 of~\cite{Thielemans:1994}.
Using the notations of this reference for the OPEs
\begin{equation}\label{eq:ope_not}
W^{s_1\, \alpha_1}(z)W^{s_2\, \alpha_2}(w) = \sum^{s_1+\frac{\alpha_1}{2}+s_2+\frac{\alpha_2}{2}}_{n=-\infty}(z-w)^{-n} [W^{s_1\, \alpha_1} W^{s_2\,\alpha_2}]_n(w)\ ,
\end{equation}
the associativity constraints can be written as
\begin{align}\label{eq:def_jacobi}
[W^{s_1\, \alpha_1}[W^{s_2\, \alpha_2}W^{s_3\, \alpha_3}]_m]_n &- (-1)^{(2s_1+\alpha_1)(2s_2+\alpha_2)}[W^{s_2\, \alpha_2}[W^{s_1\, \alpha_1}W^{s_3\, \alpha_3}]_n]_m\\
&=\sum^{s_1+\frac{\alpha_1}{2}+s_2+\frac{\alpha_2}{2}}_{l=1} \binom{n-1}{l-1}[[W^{s_1\, \alpha_1}W^{s_2\, \alpha_2}]_l W^{s_3\, \alpha_3}]_{m+n-l}\ . \notag
\end{align}
Notice that if $m,n\geq 1$, the constraints contain  information only about the singular terms in the OPEs~\eqref{eq:ope_not}.
In fact, these are the constraints which are believed to be equivalent to the Jacobi identities of the mode commutator algebra.
In the following, any reference to Jacobi identities has to be understood in the sense of eqs.~\eqref{eq:def_jacobi} for $m,n\geq 1$, which we shall compactly denote by
$[W^{s_1\,\alpha_1},W^{s_2\,\alpha_2},W^{s_3\,\alpha_3}]$.

We would like to know exactly  how strong the Jacobi identities are or, equivalently, how many structure constants
remain undetermined after solving them. Clearly, this question is related to the classification of $\mathcal{W}_\infty$-algebras.
In order to implement the ansatz~(\ref{ope22}--\ref{ope7h4}), compute composite fields and check Jacobi identities efficiently we have used the \emph{Mathematica} packages \texttt{OPEdefs} and \texttt{OPEconf} designed precisely for this task by  Thielemans.
They are described in detail in \cite{Thielemans:1991, Thielemans:1994}.\footnote{We thank K. Thielemans for providing us with the latest version of these packages.}


As can be seen from \eqref{eq:def_jacobi}, the ansatz~(\ref{ope22}--\ref{ope7h4}) enables us to compute the Jacobi identities
\begin{align}\label{eq:222}
[W^{(2)}&\; ,\;  W^{(2)}\; ,\; W^{(2)}] \ ,\\ \label{eq:227h}
[W^{(2)} &\; ,\; W^{(2)}\; ,\; W^{(\frac{7}{2})}] \ ,\\
[W^{(2)} &\; ,\; W^{(2)}\; ,\; W^{(4)}] \ , \label{eq:224}
\end{align}
without any knowledge about the OPEs $W^{(s_1)}\times W^{(s_2)}$ for total spin $s_1+s_2>\tfrac{15}{2}$.
%
%
According to our calculations, the first two Jacobi identities \eqref{eq:222} and \eqref{eq:227h} hold if and only if the structure constants appearing in (\ref{ope22}--\ref{ope7h7h}) satisfy
\begin{align}
\notag
c_{2 2}^{\frac{7}{2}} \, c_{2 \frac{7}{2}}^2 &= \tfrac{12 (5 c+6)}{c (4 c+21)} n_2  + \tfrac{3 (c-15)}{2 (5 c+6)} \left( c_{2 2}^2 \right)^{2}\ , \\ \notag
c_{2 \frac{7}{2}}^{\frac{7}{2}} &= \tfrac{4c+21}{5c+6} c_{2 2}^2\ , \\ \notag
c_{2 2}^{\frac{7}{2}} \, c_{2 \frac{7}{2}}^4 \, c_{2 4}^2 
&= -\tfrac{2 (5 c+6) (14 c-25)}{7 c^2 (10 c-7)} n_2 \, c_{2 2}^{\frac{7}{2}} \, a_{2 \frac{7}{2}}^4 
+ \tfrac{30 (5 c+6)}{7 c (2 c+29)} n_2 c_{2 2}^2 -\tfrac{26 c+177}{14 c(2 c+29)} \left( c_{2 2}^2 \right)^{2} \, c_{2 2}^{\frac{7}{2}} \, a_{2 \frac{7}{2}}^4 \\ \notag
&\quad+\tfrac{15 (c-15) (4 c+21)}{28 (2 c+29) (5 c+6)} \left( c_{2 2}^2 \right)^3  \ , 
\\ \notag
c_{2 \frac{7}{2}}^4 \, c_{2 4}^{\frac{7}{2}} &= -\tfrac{2 (4 c+21) (14 c-25)}{7 c (2 c-3) (2 c+29)} \, n_2
+ \tfrac{5 (4 c+21)}{49 c(2 c+29)} c_{2 2}^2 \, c_{2 2}^{\frac{7}{2}} \, a_{2 \frac{7}{2}}^4
-\tfrac{(4 c+21) \left(50 c^2-145 c+483\right)}{98 (2 c-3) (2 c+29) (5 c+6)}  \left( c_{2 2}^2 \right)^{2}\ , \\ \notag
%
c_{2 4}^{4} &= \tfrac{8}{7c} c_{2 2}^{\frac{7}{2}} \, a_{2 \frac{7}{2}}^4 + \tfrac{2 (25 c+84)}{7 (5 c+6)} c_{2 2}^2\ , \\ \notag
c_{2 2}^{\frac{7}{2}} \, c_{2 \frac{7}{2}}^4 \, a_{2 4}^{4} &=
-\tfrac{48 (10 c-7)}{(2 c-3) (2 c+29)} n_2 
+\tfrac{8}{7c} \left( c_{2 2}^{\frac{7}{2}}\, a_{2 \frac{7}{2}}^{4}   \right)^2  
-\tfrac{4 \left(10 c^2-51 c-1512\right)}{7 (2 c+29) (5 c+6)} c_{2 2}^2 \, c_{2 2}^{\frac{7}{2}} \,  a_{2 \frac{7}{2}}^4   \\ \notag
&\quad-\tfrac{6 c(c-15) (4 c+21) (10 c-7)}{(2 c-3) (2 c+29) (5 c+6)^2} \left( c_{2 2}^2 \right)^2\ , \\ \notag
\left( c_{2 2}^{\frac{7}{2}} \right)^2 n_{\frac{7}{2}} &= \tfrac{48 (5 c+6)}{c (4 c+21)} \left( n_2 \right)^2 + \tfrac{6 (c-15)}{5 c+6} n_2 \left( c_{2 2}^2 \right)^2\ , \\ \notag
\left( c_{2 2}^{\frac{7}{2}} \right)^2 c_{\frac{7}{2} \frac{7}{2}}^2 &= \tfrac{48}{c} n_2 c_{2 2}^2 + \tfrac{6 (c-15) (4 c+21)}{(5 c+6)^2} \left( c_{2 2}^2 \right)^3\ ,\\ \notag
c_{2 2}^{\frac{7}{2}} \, c_{\frac{7}{2} \frac{7}{2}}^{\frac{7}{2}} &= -\tfrac{36 (56 c-111)}{7 (2 c-3) (4 c+21)} n_2 
-\tfrac{9 \left(37 c^3-354 c^2-504 c
+1701\right)}{7 (2 c-3) (5 c+6)^2} \left( c_{2 2}^2 \right)^2\ , \\ \notag
c_{2 2}^{\frac{7}{2}} \, c_{\frac{7}{2} \frac{7}{2}}^{4} &= \tfrac{12(c-6)}{5c+6} c_{2 2}^2\, c_{2 \frac{7}{2}}^4 \ , \\ \notag
\left( c_{2 2}^{\frac{7}{2}} \right)^2 a_{\frac{7}{2} \frac{7}{2}}^{4} &= -\tfrac{96}{2 c-3} n_2
+ \tfrac{12 (c-6)}{5 c+6} c_{2 2}^2 \, c_{2 2}^{\frac{7}{2}} \, a_{2 \frac{7}{2}}^{4} -\tfrac{12c (c-15) (4 c+21)}{(2 c-3) (5 c+6)^2} \left( c_{2 2}^2 \right)^2 \ , \\ \notag
c_{2 2}^{\frac{7}{2}} \, c_{\frac{7}{2} \frac{7}{2}}^{\frac{11}{2}} &= - c_{2 \frac{7}{2}}^4 \, c_{2 4}^{\frac{11}{2}}\ , \\ \notag
\left( c_{2 2}^{\frac{7}{2}} \right)^2 a_{\frac{7}{2} \frac{7}{2}}^{\frac{11}{2},1} &= -\tfrac{768 c(137 c-208)}{11 (2 c-3) (2 c+29) (3 c+20)} n_2
-\tfrac{96 c(c-33) (c+11)}{(2 c+29)  (3 c+20) (5 c+6)} c_{2 2}^2 \, c_{2 2}^{\frac{7}{2}} \, a_{2 \frac{7}{2}}^{4} - c_{2 2}^{\frac{7}{2}} \, c_{2 \frac{7}{2}}^{4} \, a_{2 4}^{\frac{11}{2},1}  
\\ \notag
&\quad-\tfrac{96 c^2(c-15) (4 c+21) (137 c-208)}{11 (2 c-3) (2 c+29) (3 c+20)  (5 c+6)^2} \left( c_{2 2}^2 \right)^2  \ ,\\ \notag
c_{2 2}^{\frac{7}{2}} \, a_{\frac{7}{2} \frac{7}{2}}^{\frac{11}{2},2} &= -2 c_{2 2}^{\frac{7}{2}} \, a_{2 \frac{7}{2}}^{4} - c_{2 \frac{7}{2}}^{4} \, a_{2 4}^{\frac{11}{2},2}  
+\tfrac{81c}{11(5c+6)} c_{2 2}^2\ , \\ \notag
c_{\frac{7}{2} \frac{7}{2}}^{6} &= 0\ ,\\ \notag
\left( c_{2 2}^{\frac{7}{2}} \right)^2 \, a_{\frac{7}{2} \frac{7}{2}}^{6,1} &= \tfrac{5184c}{11  (2 c-3) (2c+61)} n_2 + \tfrac{648 c^2(c-15) (4 c+21)}{11 (2 c-3) (2c+61) (5 c+6)^2} \left( c_{2 2}^{2} \right)^2\ , \\ \notag 
c_{2 2}^{\frac{7}{2}} \, a_{\frac{7}{2} \frac{7}{2}}^{6,2} &= - \tfrac{378c}{11(5c+6)} c_{2 2}^2\ , \\
a_{\frac{7}{2} \frac{7}{2}}^{6,3} &= a_{\frac{7}{2} \frac{7}{2}}^{6,4} = a_{\frac{7}{2} \frac{7}{2}}^{\frac{13}{2},1} = 
a_{\frac{7}{2} \frac{7}{2}}^{\frac{13}{2},2} = 0\ .
\label{eq:jacobi} 
\end{align}
Observe that $c_{\frac{7}{2}\frac{7}{2}}^6=0$ and, as a consequence, the field $W^{(6)}$ together with its descendants cannot appear on the r.h.s.\ of \eqref{ope7h7h}.
For this reason, the ans\"atze~\eqref{ope211h} and \eqref{ope7h4} are actually enough not only  to check the Jacobi identity \eqref{eq:224},
but also
\begin{equation}\label{eq:27h7h}
[W^{(2)}\; ,\; W^{(\frac{7}{2})}\; ,\; W^{(\frac{7}{2})}]\ .
\end{equation}
The solution to the Jacobi identities \eqref{eq:224} and \eqref{eq:27h7h} is rather lengthy and
we shall write it down only after conveniently fixing some renormalization and redefinition freedom of the generators $W^{(s)}$.
This will also make the structure of eqs.~\eqref{eq:jacobi} more transparent.

First, notice that one can exploit the renormalization freedom of the generators $W^{(2)}$, $W^{(\frac{7}{2})}$, $W^{(4)}$, $W^{(\frac{11}{2})}$ and $W^{(6)}$ to fix  the  structure constants
\begin{equation}\label{eq:norm_str}
c_{22}^2 = c_{22}^{\frac{7}{2}}= c_{2\frac{7}{2}}^4=c_{24}^{\frac{11}{2}}=c_{2\frac{11}{2}}^6=1\ .
\end{equation}
In this normalization, the  central terms $n_s$ in \eqref{eq:ns} cannot be chosen arbitrarily.
They play  the role of non-trivial structure constants and must be computed from the Jacobi identities.
The reason we prefer to absorb the renormalization freedom in the structure constants~\eqref{eq:norm_str} rather then in $n_s$
is that in the latter case a residual sign flip symmetry $W^{(s)}\mapsto -W^{(s)}$ survives.
In order to avoid dealing with ambiguities arising from this discrete symmetry later, we prefer to choose  a more convenient normalization from the very start.

Second, notice that eqs.~(\ref{ope22}--\ref{ope7h4})  define the $\mathcal{W}_\infty$-algebra structure in a basis of $\mathcal{N}=1$ \emph{primary} generators.
On the other hand, the set $\{W^{(s)}\}$ is not the only such basis.
For $s\geq4$ one has the freedom to shift the generators $W^{(s)}$ by $\mathcal{N}=1$ primary composite fields.
Thus, a given $\mathcal{W}_\infty$-algebra structure can be represented in many different primary bases
\begin{align*}
&\hat{W}^{(2)}=W^{(2)}\ ,\quad  \hat{W}^{(\frac{7}{2})}=W^{(\frac{7}{2})}\ ,\quad
\hat{W}^{(4)} = W^{(4)}+\alpha^{(4)}A^{(4)}\ ,\\
&\hat{W}^{(\frac{11}{2})}=W^{(\frac{11}{2})}+\alpha^{(\frac{11}{2},1)}A^{(\frac{11}{2},1)}+\alpha^{(\frac{11}{2},2)}A^{(\frac{11}{2},2)}\ ,\quad \cdots
\end{align*}
by OPEs of the same general  form~(\ref{ope22}--\ref{ope7h4}).  As  mentioned in \cite{Candu:2012ne}, this redefinition freedom for the generators implies some consistency conditions for the Jacobi identities \eqref{eq:jacobi}.
We shall use it in our favor to fix a few more structure constants
\begin{equation}\label{eq:redef}
a_{2\frac{7}{2}}^4= a_{24}^{\frac{11}{2},1}= a_{24}^{\frac{11}{2},2}=
a_{2\frac{11}{2}}^{6,1}= \cdots = a_{2\frac{11}{2}}^{6,4}=0\ .
\end{equation}

The Jacobi identities~\eqref{eq:jacobi} simplify considerably after  the renormalization and redefinition freedom of the $\mathcal{W}_\infty$-generators is fixed as in eqs.~(\ref{eq:norm_str}, \ref{eq:redef})
\begin{align} \notag
c_{2 \frac{7}{2}}^2 &= \tfrac{3 (c-15)}{2 (5 c+6)} + \tfrac{12 (5 c+6)}{c (4 c+21)} n_2 \ , &
c_{2 \frac{7}{2}}^{\frac{7}{2}} &= \tfrac{4 c+21}{5 c+6} \ , \\ \notag
c_{2 4}^2 
&= \tfrac{15 (c-15) (4 c+21)}{28 (2 c+29) (5 c+6)} + \tfrac{30 (5 c+6)}{7 c (2 c+29)} n_2 \ , &
c_{2 4}^{\frac{7}{2}} &= -\tfrac{(4 c+21) (50 c^2-145 c+483 )}{98 (2 c-3) (2 c+29) (5 c+6)} 
-\tfrac{2 (4 c+21) (14 c-25)}{7 c (2 c-3) (2 c+29)} n_2 \ , \\ \notag
c_{2 4}^{4} &= \tfrac{2 (25 c+84)}{7 (5 c+6)} \ , &
a_{2 4}^{4} &= -\tfrac{6 c(c-15) (4 c+21) (10 c-7)}{(2 c-3) (2 c+29) (5 c+6)^2} -\tfrac{48 (10 c-7)}{(2 c-3) (2 c+29)} n_2 \ , \\ \notag
n_{\frac{7}{2}} &= \tfrac{6 (c-15)}{5 c+6} n_2 + \tfrac{48 (5 c+6)}{c (4 c+21)} n_2^2 \ , &
c_{\frac{7}{2} \frac{7}{2}}^2 &= \mathrlap{\tfrac{6 (c-15) (4 c+21)}{(5 c+6)^2} + \tfrac{48}{c} n_2 \ ,} \\ \notag
c_{\frac{7}{2} \frac{7}{2}}^{\frac{7}{2}} &= \mathrlap{-\tfrac{9 (37 c^3-354 c^2-504 c+1701)}{7 (2 c-3) (5 c+6)^2}  -\tfrac{36 (56 c-111)}{7 (2 c-3) 
(4 c+21)} n_2 \ ,} 
\\ \notag
c_{\frac{7}{2} \frac{7}{2}}^{4} &= \tfrac{12 (c-6)}{5 c+6} \ , &
a_{\frac{7}{2} \frac{7}{2}}^4 &= \mathrlap{-\tfrac{12 c(c-15) (4 c+21)}{(2 c-3) (5 c+6)^2} -\tfrac{96}{2 c-3} n_2 \ ,} \\ \notag
c_{\frac{7}{2} \frac{7}{2}}^{\frac{11}{2}} &= -1 \ ,\\ \notag
a_{\frac{7}{2} \frac{7}{2}}^{\frac{11}{2},1} &= \mathrlap{-{}\tfrac{96 c^2(c-15) (4 c+21) (137 c-208)}{11 (2 c-3) (2 c+29) (3 c+20) (5
   c+6)^2} -\tfrac{768 c(137 c-208)}{11 (2 c-3) (2 c+29) (3 c+20)} n_2 \ ,} \\ \notag
a_{\frac{7}{2} \frac{7}{2}}^{\frac{11}{2},2} &= \tfrac{81c}{11 (5 c+6)} \ , &
c_{\frac{7}{2} \frac{7}{2}}^{6} &= 0 \ , \\ \notag
a_{\frac{7}{2} \frac{7}{2}}^{6,1} &= \mathrlap{\tfrac{648 c^2(c-15) (4c+21)}{11 (2 c-3) (2 c+61) (5 c+6)^2} + \tfrac{5184c}{11 (2 c-3) (2 c+61)} n_2 \ ,} \\ 
a_{\frac{7}{2} \frac{7}{2}}^{6,2} &= - \tfrac{378c}{11 (5 c+6)} \ , &
a_{\frac{7}{2} \frac{7}{2}}^{6,3} &= a_{\frac{7}{2} \frac{7}{2}}^{6,4} = a_{\frac{7}{2} \frac{7}{2}}^{\frac{13}{2},1} = 
a_{\frac{7}{2} \frac{7}{2}}^{\frac{13}{2},2} = 0 \ .
\label{jacobi_simple}
\end{align}
We see explicitly that these equations determine \emph{all} the structure constants in the OPEs (\ref{ope22}--\ref{ope7h7h})
\emph{uniquely} in terms of the central charge $c$ and the central term 
$n_2$.
 This is of course only true for generic values of $c$.
Thus, it seems at this point that the $\mathcal{W}_\infty$-algebra depends only on two fundamental parameters --- $c$ and $n_2$.

We can now write down in a reasonably compact form also the unique solution to the Jacobi identities \eqref{eq:224}
and \eqref{eq:27h7h} 
\begin{align}
\notag
c_{2 \frac{11}{2}}^2 &= 
\mathrlap{\tfrac{3 (c-15) (4 c+21) (200 c^4-1898 c^3-46441
   c^2+66031 c-27836)}{4 (2 c-3) (2 c+29) (2 c+53) (3 c+20) (5 c+6) (28 c-17)}} \\ \notag
&\quad \mathrlap{{}+\tfrac{3 (11568 c^6-237332 c^5-1219712 c^4+10601401 c^3-2032733
   c^2-55167798 c+31196808)}{c (2 c-3) (2 c+29) (2 c+53) (3 c+20) (5 c+6) (28 c-17)} n_2} \\ \notag
&\quad \mathrlap{{}+\tfrac{24 (5 c+6) (14 c-25) (c^2-94 c-248 )}{c^2 (2 c-3) (2 c+29) (2 c+53) (3 c+20)} n_2^2 \ ,}
\\ \notag
c_{2 \frac{11}{2}}^{\frac{7}{2}} &= \mathrlap{\tfrac{3 (4 c+21) (400 c^5+5164 c^4-35500 c^3-297643 c^2+1018362 c-565614)}{2 (2 c-3) (2 c+29) (2 c+53) 
(3 c+20) (5 c+6) (28 c-17)}}
\\ \notag
&\quad \mathrlap{{}+ \tfrac{3 (4 c+21) (8488 c^5+160420 c^4+352934 c^3-2400661 c^2+2335062 c-572040)}{c (2 c-3) (2 c+29) (2 c+53) (3 c+20) (5 c+6) (28
   c-17)} n_2 \ ,}\\ \notag 
c_{2 \frac{11}{2}}^{4} &= \mathrlap{\tfrac{3 (100 c^5+1646 c^4-7134 c^3-273680 c^2-521053 c+2214744)}{2 (2 c-3) (2 c+29) (2 c+53) (3 c+20) (5 c+6)} +
\tfrac{12 (90 c^4+1881 c^3+13678 c^2-316 c-66824)}{c (2 c-3) (2 c+29) (2 c+53) (3 c+20)} n_2 \ ,}
\\ \notag
a_{2 \frac{11}{2}}^{4} &= \mathrlap{-\tfrac{9 c(c-15) (4 c+21) (10 c-7) (6 c^2-2425
   c+1524)}{4 (2 c-3) (2 c+29) (3 c+20) (5 c+6)^2 (28 c-17)} -\tfrac{18 (10 c-7) (6 c^2-2425 c+1524)}{(2 c-3) (2 c+29) (3 c+20) (28 c-17)} n_2 \ ,} \\ \notag
c_{2 \frac{11}{2}}^{\frac{11}{2}} &= \mathrlap{\tfrac{(c+12) (20 c+61)}{(3 c+20) (5 c+6)} \ ,} \\ \notag 
a_{2 \frac{11}{2}}^{\frac{11}{2} ,1} &= \mathrlap{-\tfrac{144 c^2(c-15) (c+11) (4 c+21) (10 c-7)
   (173 c^2-3569 c+2208)}{(2 c-3) (2 c+29) (3 c+20)^2 (5 c+6)^3 (28 c-17)} 
-\tfrac{1152 c(c+11) (10 c-7) (173 c^2-3569 c+2208)}{(2 c-3) (2 c+29) (3 c+20)^2 (5 c+6) (28 c-17)} n_2 \ ,}
\\ \notag
a_{2 \frac{11}{2}}^{\frac{11}{2},2} &= \mathrlap{-\tfrac{3 c(2060 c^5+26898 c^4-51828 c^3-4090735   c^2-15570789 c+13204800)}{(2 c-3) (2 c+29) (2 c+53) 
(3 c+20) (5 c+6)^2} -\tfrac{168 (62 c^3+1083 c^2+7590 c-6436)}{(2 c-3) (2 c+29) (2 c+53) (3 c+20)} n_2 \ ,}\\ \notag
c_{\frac{7}{2} 4}^2 &= \mathrlap{-\tfrac{3 (c-15) (4 c+21) (50 c^2-145 c+483)}{28 (2 c-3) (2 c+29) (5 c+6)^2}-\tfrac{3
   (892 c^3-5372 c^2-20955 c+60921)}{7 c (2 c-3) (2 c+29) (5 c+6)} n_2 -\tfrac{24 (5 c+6) (14 c-25)}{c^2 (2 c-3) (2 c+29)} n_2^2 \ ,} \\ \notag
c_{\frac{7}{2} 4}^{\frac{7}{2}} &= \mathrlap{\tfrac{3 (4 c+21) (50 c^3-485 c^2+1743 c+252)}{28 (2 c-3) (2
   c+29) (5 c+6)^2} + \tfrac{3 (4 c+21) (98 c^2-863 c+930)}{7 c (2 c-3) (2 c+29) (5 c+6)} n_2 \ ,} \\ \notag
c_{\frac{7}{2} 4}^{4} &= \mathrlap{-\tfrac{3 (600 c^4-14 c^3-52955 c^2-58431 c+270144)}{14 (2 c-3)(2 c+29) (5 c+6)^2} 
-\tfrac{24 (30 c^3+197 c^2-679 c+123)}{c (2 c-3) (2 c+29) (4 c+21)} n_2 \ ,} \\ \notag
a_{\frac{7}{2} 4}^{4} &= \mathrlap{-\tfrac{81c (c-15) (4 c+21) (10 c-7)}{14 (2 c-3) (5 c+6)^3} -\tfrac{324 (10 c-7)}{7 (2 c-3) (5 c+6)} n_2 \ ,}\\ \notag
c_{\frac{7}{2} 4}^{\frac{11}{2}} &= \tfrac{9 (2 c+21)}{7 (5 c+6)} \ , &
a_{\frac{7}{2} 4}^{\frac{11}{2} ,1} &= \tfrac{2592 c^2(c-15) (c+11) (4 c+21) (10 c-7)}{7 (2 c-3) (2 c+29) (3 c+20) (5 c+6)^3}
+\tfrac{20736 c(c+11) (10 c-7)}{7 (2 c-3) (2 c+29) (3 c+20) (5 c+6)} n_2 \ , \\ \notag
a_{\frac{7}{2} 4}^{\frac{11}{2},2} &= \mathrlap{-\tfrac{6 c(530 c^3-3351 c^2-18594 c+18333 )}{7 (2 c-3) (2 c+29) (5 c+6)^2}
-\tfrac{24 (34 c-39)}{(2 c-3) (2 c+29)} n_2 \ ,} \\ \notag
c_{\frac{7}{2} 4}^{6} &= -1 \ , \\ \notag
a_{\frac{7}{2} 4}^{6,1} &= 
\mathrlap{\tfrac{2430 c^2(c-15) (c+11) (4 c+21) (10 c-7) (14 c+11)}{77 (2 c-3) (2
   c+29)^2 (2 c+61) (5 c+6)^2 (28 c-17)} +
\tfrac{19440 c(c+11) (10 c-7) (14 c+11)}{77 (2 c-3) (2 c+29)^2 (2 c+61) (28 c-17)} n_2 \ ,} \\ \notag
a_{\frac{7}{2} 4}^{6,2} &= \mathrlap{-\tfrac{3 c(19 c+104) (50 c^2-145 c+483)}{88 (2 c-3) (2 c+29) (2 c+53) (5 c+6)}
-\tfrac{21 (14 c-25) (19 c+104)}{22 (2 c-3) (2 c+29) (2 c+53)} n_2 \ ,} \\ \notag
a_{\frac{7}{2} 4}^{6,3} &= \tfrac{4 c(c+48)}{(3 c+20) (5 c+6)} \ , &
a_{\frac{7}{2} 4}^{6,4} &= \tfrac{42 c^2(c-15) (4 c+21) (10 c-7)}{(2 c-3) (2 c+29) (3 c+20) (5 c+6)^2}+\tfrac{336 c(10 c-7)}{(2 c-3) (2 c+29) 
(3 c+20)} n_2 \ , \\ \notag
a_{\frac{7}{2} 4}^{\frac{13}{2},1}  &=  \mathrlap{\tfrac{c(990 c^3-21283 c^2-168837 c+108990)}{56 (2 c-3) (2 c+29) (5 c+6)^2}
+\tfrac{90 c+13}{2 (2 c-3) (2 c+29)} n_2 \ ,} \\ \notag
a_{\frac{7}{2} 4}^{\frac{13}{2},2} &= \tfrac{18c}{5 c+6} \ , &
a_{\frac{7}{2} 4}^{7,1} &=  \tfrac{9 c^2\left(530 c^3-3351 c^2-18594 c+18333\right)}{(c+11) (2 c-3) (2 c+29) (5 c+6)^2}
+\tfrac{252 c(34 c-39)}{(c+11) (2 c-3) (2 c+29)} n_2 \ , \\ \notag 
a_{\frac{7}{2} 4}^{7,2} &= \tfrac{27c}{13 (5 c+6)} \ , \\ \notag
a_{2 \frac{11}{2}}^{\frac{13}{2},1} &= \mathrlap{-\tfrac{c(21690 c^5+300777 c^4-4114852 c^3-36985693
   c^2-45629394 c+52993080)}{56 (2 c-3) (2 c+29) (2 c+53) (3 c+20) (5 c+6)^2}} \\ \notag
&\mathrlap{\quad{}-\tfrac{1422 c^3+30045 c^2+44995 c-52988}{2 (2 c-3) (2 c+29) (2 c+53) (3 c+20)} n_2  \ ,} \\ \notag
a_{2 \frac{11}{2}}^{\frac{13}{2},2} &= \mathrlap{\tfrac{c(23 c-12)}{(3 c+20) (5 c+6)} \ ,} \\ \notag
a_{2 \frac{11}{2}}^{7,1} & = 
\mathrlap{{}-\tfrac{9 c^2\left(5820 c^5+13306 c^4+512764 c^3+5029629
   c^2+6813351 c-5964840\right)}{2 (c+11) (2 c-3) (2 c+29) (2 c+53) (3 c+20) (5 c+6)^2}} \\ \notag
&\mathrlap{\quad{}-\tfrac{252 c\left(150 c^3-473 c^2-4490 c+468\right)}{(c+11) (2 c-3) (2 c+29) (2 c+53) (3 c+20)} n_2 \ ,} \\
a_{2 \frac{11}{2}}^{7,2} &= \mathrlap{\tfrac{9 c(23 c-12)}{13 (3 c+20) (5 c+6)} \ .}
\label{jacobi_higher}
\end{align}
Again, we see that $n_2$ remains unconstrained, whereas 
all structure constants appearing in the two OPEs (\ref{ope211h}--\ref{ope7h4}) 
are uniquely determined in terms of $c$ and $n_2$.
This supports our claim that $c$ and $n_2$ are the only two parameters of $\mathcal{W}_\infty$.

Expecting the  higher level Jacobi identities to have a similar structure, we conjecture  that the Jacobi identities determine  all the structure constants of the $\mathcal{W}_\infty$-algebra uniquely in terms of $c$ and $n_2$ up to the renormalization and redefinition
freedom of its generators.
In a convenient choice of basis, extending \eqref{eq:norm_str} and \eqref{eq:redef}, these should be polynomials in $n_2$, the   coefficients being rational functions of $c$.
This is analogous to what was found for the three other infinitely generated $\mathcal{W}$-algebras relevant for the original
minimal model holography~\cite{Gaberdiel:2012}, and its $\mathcal{N}=2$ \cite{Candu:2012tr} and even spin \cite{Candu:2012ne} generalizations.

Whenever we want to emphasize the dependence of $\mathcal{W}_\infty$ on $c$ and $n_2$ we shall write $\mathcal{W}^{n_2,c}_\infty$.\footnote{Notice that if 
one restores an arbitrary normalization for $W^{(2)}$, then the parameter $n_2$ appearing in eqs.~(\ref{jacobi_simple}, \ref{jacobi_higher})
must be identified with $n_2\mapsto n_2\,/ \left(c_{22}^2\right)^2$.
It is therefore clear that one can trade $n_2$ for $\gamma:=\left(c_{22}^2\right)^2$ simply by changing the normalization conventions.
In fact, this latter choice was preferred in the previous works \cite{Gaberdiel:2012, Candu:2012tr, Candu:2012ne} and we hope that our new conventions will not be a source of
confusion.
\label{ft}}
Next, we study the large $c$ limit of $\mathcal{W}^{n_2,c}_\infty$ and trace the origin of the continuous parameter $n_2$ back to the parameter $\mu$ of the Lie superalgebra $\wg{\mu}$ mentioned in the introduction.


\subsection{Wedge algebra}
\label{sec:wedge}


The Lie algebra of the wedge modes --- called the wedge algebra ---
%
%
%
captures much of the information encoded in the  $\mathcal{W}$-algebra, for instance the spectrum of generators, but it is much simpler to deal with.
In fact, if one assumes that the  $\mathcal{W}$-algebra is of a Drinfel'd-Sokolov (DS) type,
then the wedge algebra identifies it uniquely \cite{Bowcock:1991zk}.


Let us recall the definition of the wedge algebra. A mode $W_n$ of a quasiprimary field
\begin{equation*}
W(z) = \sum_{n\in \mathbb{Z}-s}  W_nz^{-n-s} 
\end{equation*}
is said to lie  within the wedge if $|n|<s$, where $s$ is the spin of $W(z)$.
It was shown in~\cite{Bowcock:1991zk} that the commutation relations of the wedge modes of the generators of a $\mathcal{W}$-algebra
 close on themselves in the  $c\to\infty$ limit.
The assumptions under which this was derived are: (i) the $\mathcal{W}$-algebra exists for arbitrarily large values of $c$, (ii) it has a well-defined classical limit and (iii) its
generators can be normalized in such a way that the coefficients of all generators appearing in the singular part of the OPE of any two 
generators are of order $\mathcal{O}(c^0)$.
%

Let us now look at the wedge algebra limit of $\mathcal{W}_\infty$.
First, we remove the generator redefinition freedom  using eqs.~(\ref{eq:norm_str}, \ref{eq:redef}).
Next, notice that all structure constants $c_{ss'}^{s''}$ in eqs.~(\ref{jacobi_simple}, \ref{jacobi_higher}) remain of order $\mathcal{O}(c^0)$ in the $c\to\infty$ limit if we keep the ratio $n_2/c$ finite.
Thus, the conditions (i) and (iii) for the existence of the wedge algebra  are  satisfied and
we shall see in sec.~\ref{sec:classical} that condition (ii) is also satisfied, at least up to the level  we have constructed the $\mathcal{W}_\infty$-algebra.
In the following we shall assume that these conditions  hold  in general.

Now, recall that   global $\mathcal{N}=1$ conformal symmetries $\{L_0, L_{\pm 1}, G_{\pm 1/2}\}$ form an $\mathfrak{osp}(1|2)$ Lie superalgebra.
Under their action, the wedge modes of the supermultiplet $W^{(s)}$
transform in an irreducible representation of $\mathfrak{osp}(1|2)$, which decomposes under the action of the $\mathfrak{sl}(2)$ subalgebra $\{L_0, L_{\pm 1}\}$
into two representations of spin $j=s-1$ and $j=s-\tfrac{1}{2}$ corresponding to the wedge modes of $W^{s\,0}$ and $W^{s\,1}$. We shall denote it by $\langle s,s+\frac{1}{2}\rangle$. 
Thus, the wedge algebra of $\mathcal{W}_\infty$, which we denote by $\mathcal{W}^\wedge_\infty$, must be a Lie superalgebra naturally equipped with an $\mathfrak{osp}(1|2)$ embedding such that w.r.t.\ the latter it decomposes as
\begin{equation}\label{eq:wedgedef}
\mathcal{W}_\infty^\wedge \big\vert_{\mathfrak{osp}(1|2)}
\simeq \bigoplus_{s\in 2\mathbb{N}-\frac{1}{2}}\langle s,s+\tfrac{1}{2}\rangle \oplus  \bigoplus_{s\in 2\mathbb{N}}\langle s,s+\tfrac{1}{2}\rangle\ .
\end{equation}
%
%
In order to identify it, one must turn the OPEs (\ref{ope22}--\ref{ope7h4}) into commutators, following for instance \cite{Bowcock:1990ku},
restrict to the wedge and then take the $c\to\infty$ limit, while keeping the ratio
$n_2/c$ finite.
The resulting commutation relations will be of the form
\begin{equation}\label{eq:comWedge}
[W^{s\, \alpha}_m,W^{s'\, \alpha'}_{m'}] = \sum_{s''} P^{s+\frac{\alpha}{2}\,s'+\frac{\alpha'}{2}}_{s''+\frac{\alpha''}{2}}(m,m')\, h_{ss'}^{s''}(\alpha,\alpha') W^{s''\, \alpha''}_{m+m'}\ ,
\end{equation}
where we have set $W^{\frac{3}{2}\, 0}:= \frac{1}{2}\, G$ and $W^{\frac{3}{2}\, 1}:= T$.
All  non-trivial information is contained in the structure constants $h_{ss'}^{s''}(\alpha,\alpha')$ defined as
\begin{equation}
c_{ss'}^{s''}(\alpha,\alpha') = h_{ss'}^{s''}(\alpha,\alpha')+\mathcal{O}(c^{-1})\ .
\label{eq:corr1}
\end{equation}
Here $c_{ss'}^{\frac{3}{2}}(\alpha,\alpha')$ denotes the coefficient of the quasiprimary $W^{\frac{3}{2}\,\alpha''}$ in the OPE $W^{s\,\alpha}(z) W^{s'\,\alpha'}(w)$.
The latter is clearly non-zero only when $s=s'$ and it is easy to see that
\begin{equation}
c_{ss}^{\frac{3}{2}}:=c_{ss}^{\frac{3}{2}}(0,0)=\frac{2s n_s}{c}\ ,
\label{eq:corr1_mod}
\end{equation}
while the remaining components $c_{ss}^{\frac{3}{2}}(\alpha,\alpha')$ can be computed from eq.~\eqref{eq:rel1}.
The polynomials $P^{ss'}_{s''}(m,m')$ contain the mode dependence of the commutators
and are entirely fixed by global conformal symmetry
\begin{equation}\label{eq:uni_pol}
P^{j+1\ j'+1}_{j''+1}(m,m') :=\sum_{r=0}^{j+j'-j''}\binom{j+m}{j+j'-j''-r}\frac{ (-1)^r (j-j'+j''+1)_{(r)}(j''+m+m'+1)_{(r)} }{r! (2j''+2)_{(r)}} \ ,
\end{equation}
where $(a)_{(n)}:=\Gamma(a+n)/\Gamma(a)$ is the Pochhammer symbol.

Let us introduce a more compact, OPE like notation for the commutators in \eqref{eq:comWedge} 
\be\label{eq:compact_comm_shss}
 [W^{(s)}, W^{(s')}]  = \sum_{s''} h_{ss'}^{s''}\, W^{(s'')} \ ,
\ee
where  $h_{ss'}^{s''}$ is the $c\to\infty$ limit of $c_{ss'}^{s''}$ defined in eq.~(\ref{eq:def_c}, \ref{eq:corr1_mod}).
Now, if we identify 
\begin{equation}\label{eq:wedge_match}
\frac{n_2}{c} = -\frac{(\mu -2) (\mu +1)}{8 (2 \mu-1)^2}+\mathcal{O}(c^{-1})\ ,
\end{equation}
then the wedge algebra commutators extracted from the OPEs~(\ref{ope22}--\ref{ope7h7h}) with the help of eq.~\eqref{jacobi_simple} take the following explicit
form
\begin{align} \label{eq:hscom1}
[W^{(2)}, W^{(2)}] & = -\tfrac{(\mu -2) (\mu +1)}{2 (2 \mu-1)^2} W^{(\frac{3}{2})}  + W^{(2)} + W^{(\frac{7}{2})}\ , \\ \notag
 [W^{(2)}, W^{(\frac{7}{2})}]  & = -\tfrac{27 (\mu -3) (\mu +2)}{40 (2 \mu-1)^{2}} W^{(2)} 
+\tfrac{4}{5} W^{(\frac{7}{2})} +  W^{(4)}  \ ,\\  \notag
 [W^{(2)}, W^{(4)}]  & = \tfrac{9 (\mu -4) (\mu +3)}{98 (2 \mu-1)^2} 
W^{(\frac{7}{2})}
+\tfrac{10}{7} W^{(4)} + W^{(\frac{11}{2})} \ , \\ \notag
 [W^{(\frac{7}{2})}, W^{(\frac{7}{2})}]  & = \tfrac{189 (\mu -3) (\mu -2) (\mu +1) (\mu +2)}{80 (2 \mu -1)^4}
W^{(\frac{3}{2})} - \tfrac{54 (\mu -3) (\mu +2)}{25 (2 \mu -1)^2} W^{(2)} 
+ \tfrac{81 \left(3 \mu ^2-3 \mu -43\right)}{350 (2 \mu -1)^2} W^{(\frac{7}{2})} \\ \notag
&\quad+\tfrac{12}{5} W^{(4)} - W^{(\frac{11}{2})} \ . 
\end{align}
%
These commutation relations match precisely with the commutation relations of 
$\mathfrak{shs}^\sigma[\mu]$ --- a subalgebra of the higher spin superalgebra $\mathfrak{shs}[\mu]$ fixed by a $\mathbb{Z}_2$ automorphism $\sigma$.
The exact definition of $\mathfrak{shs}[\mu]$ and the explicit form of the automorphism $\sigma$ can be found in app.~\ref{shssigma}.
In there we give the relation between  $W^{s\,\alpha}_m$ and the basis in which the commutation relations of $\mathfrak{shs}^\sigma[\mu]$ are known explicitly.
Eq.~\eqref{eq:hscom1} strongly suggests that the wedge algebra of $\mathcal{W}_\infty^{n_2,c}$ is isomorphic to $\mathfrak{shs}^\sigma[\mu]$, where $\mu$ is determined by $n_2$ and $c$ according to eq.~\eqref{eq:wedge_match}.


\subsection{Classical analysis}
\label{sec:classical}


The result of the previous section suggests that all $\mathcal{W}_\infty$-algebras can be realized as quantum Drinfel'd-Sokolov reductions of $\wg{\mu}$.
Unfortunately, making sense of this statement is quite problematic at the moment because the standard
quantum DS reduction procedure gets plagued with divergences when applied to the case of an infinite dimensional infinite rank Lie superalgebra such as $\mathfrak{shs}^\sigma[\mu]$.\footnote{However,
it is tempting to believe that a generalization of the  quantum DS reduction overcoming these difficulties exists, especially in the supersymmetric case, 
see \cite{Fradkin:1990qk}.}
On the other hand, the algebra
\begin{equation}\label{eq:wcl_DS_def}
\mathcal{W}^{cl}_\infty[\mu]\; :=\; \text{classical DS reduction of } \mathfrak{shs}^\sigma[\mu]\ 
\end{equation}
is well-defined and coincides with the asymptotic symmetry algebra of the $\mathfrak{shs}^\sigma[\mu]$ Chern-Simons theory
subject to AdS$_3$ boundary conditions. This is because, technically, asymptotic symmetries are always constructed
as the Drinfel'd-Sokolov reduction of the gauge algebra of the CS theory, see~\cite{Henneaux:2010xg, Campoleoni:2010zq, Campoleoni:2011hg, Gaberdiel:2011wb, Henneaux:2012ny, Hanaki:2012yf}.
The aim of this section is to show that \emph{any} classical $\mathcal{W}_\infty$-algebra generated by the spins~\eqref{eq:Winf_spec} is isomorphic to $\mathcal{W}_\infty^{cl}[\mu]$
for some value of $\mu$, provided that the classification conjecture of sec.~\ref{sec:classif} and a similar classification conjecture in the classical setting (see bellow) hold.
For the moment it is only clear that these algebras have the same spectrum of generators, compare eq.~\eqref{eq:wedgedef} with (\ref{eq:wedge}).

As a first step to achieve our goal, let us check whether  the most general classical $\mathcal{W}_\infty$-algebra has the same number of parameters as $\mathcal{W}_\infty^{cl}[\mu]$
by repeating the analysis of  sec.~\ref{sec:constr} and \ref{sec:classif}  in the classical setting.
We shall mark with a tilde the classical counterparts of all operators and structure constants introduced in the quantum case.
The starting point is to formulate the most general ansatz for the Poisson brackets between the generators
\begin{equation}
\{\tilde{W}^{s_1\, \alpha_1}(x_1),\tilde{W}^{s_2\, \alpha_2}(x_2)\} = \sum^{s_1+\frac{\alpha_1}{2}+s_2+\frac{\alpha_2}{2}}_{n=1}
\{\tilde{W}^{s_1\, \alpha_1} \tilde{W}^{s_2\,\alpha_2}\}_n(x_2)
\frac{(-1)^{n-1}\partial_{x_1}^{n-1}\delta(x_1-x_2)}{(n-1)!}\ ,\label{eq:cl_ansatz}
\end{equation}
where we have borrowed the notations of \cite{Thielemans:1994}.
The superconformal symmetry determines  the coefficients of all the fields on the r.h.s.\ in terms of the coefficients of $\mathcal{N}=1$ primaries only.
Explicit expressions for the classical $\mathcal{N}=1$ primaries up to spin $s=7$ are given  in app.~\ref{sec:composite} by taking the classical limit of their quantum analogs.
Thus,  the most general ansatz for the Poisson brackets \eqref{eq:cl_ansatz} with $s_1+s_2\leq \tfrac{15}{2}$ will have
the same structure as the ansatz~(\ref{ope22}--\ref{ope7h4}) for the quantum OPEs --- there is a one to one correspondence between classical and quantum $\mathcal{N}=1$ primaries and their structure constants.
The next step is to compute  the Jacobi identities
\begin{equation*}\label{eq:}
\{\tilde{W}^{s_1\, \alpha_1}(x_1),\{\tilde{W}^{s_2\, \alpha_2}(x_2),\tilde{W}^{s_3\, \alpha_3}(x_3)\}\}+\text{ grad. cycl. } = 0\ ,
\end{equation*}
and classify their solutions. These can be rewritten in a form
\begin{align}\label{eq:def_jacobi_cl}
\{\tilde{W}^{s_1\, \alpha_1}\{\tilde{W}^{s_2\, \alpha_2}\tilde{W}^{s_3\, \alpha_3}\}_m\}_n &- (-1)^{(2s_1+\alpha_1)(2s_2+\alpha_2)}\{\tilde{W}^{s_2\, \alpha_2}\{\tilde{W}^{s_1\, \alpha_1}\tilde{W}^{s_3\, \alpha_3}\}_n\}_m\\
&=\sum^{s_1+\frac{\alpha_1}{2}+s_2+\frac{\alpha_2}{2}}_{l=1} \binom{n-1}{l-1}\{\{\tilde{W}^{s_1\, \alpha_1}\tilde{W}^{s_2\, \alpha_2}\}_l \tilde{W}^{s_3\, \alpha_3}\}_{m+n-l}\ , \notag
\end{align}
which is perfectly analogous to the quantum case~\eqref{eq:def_jacobi}.
We shall compactly denote the set of Jacobi identities~\eqref{eq:def_jacobi_cl} for various $\alpha$'s and $m,n\geq 1$ by $\{\tilde{W}^{(s_1)},\tilde{W}^{(s_2)},\tilde{W}^{(s_3)}\}$.

To carry  out these calculations efficiently we used  the \emph{Mathematica} package \texttt{OPEdefs}, which can be switched to
``classical OPEs''.\footnote{The command for the switch is \texttt{SetOPEOptions[OPEMethod, ClassicalOPEs]}. The package \texttt{OPEconf} does not support this option.
In the quantum case, its functionalities make the analysis much simpler.
We did actually use it  in the classical analysis after adding a few lines to support the option \texttt{ClassicalOPEs}.}
We can summarize the results as follows: all relations between classical structure constants imposed by  the classical  Jacobi identities
\begin{equation*}
\{\tilde{W}^{(2)},\tilde{W}^{(2)},\tilde{W}^{(2)}\}\ ,\quad \{\tilde{W}^{(2)},\tilde{W}^{(2)},\tilde{W}^{(\frac{7}{2})}\}\ ,\quad \{\tilde{W}^{(2)},\tilde{W}^{(2)},\tilde{W}^{(4)}\}\ ,\quad 
 \{\tilde{W}^{(2)},\tilde{W}^{(\frac{7}{2})},\tilde{W}^{(\frac{7}{2})}\}
\end{equation*}
can be obtained by taking the $c\to\infty$ limit of eqs.~(\ref{eq:jacobi}, \ref{jacobi_simple}, \ref{jacobi_higher}) with $n_2/c$ fixed and then decorating the result with tildes.
Consequently, the Jacobi identities determine all the structure constants of the classical $\mathcal{W}_\infty$-algebra, at least up to the level 
that we have considered, uniquely in terms of
$\tilde{n}_2$ and $\tilde{c}$  if we remove the freedom of redefining the generators in the same way as we did in the quantum case~(\ref{eq:norm_str}, \ref{eq:redef}).
Put differently, the classical Jacobi identities are as constraining as the quantum ones, i.e.\ no anomalies appear.
Therefore, in analogy to the quantum case, we conjecture that the space of classical $\mathcal{W}_\infty$-algebras is parametrized by $\tilde{n}_2$ and $\tilde{c}$.

Let us now establish the relation between $\tilde{n}_2$ and the parameter $\mu$ of $\mathcal{W}_\infty^{cl}[\mu]$.
It was noticed in \cite{Bowcock:1991zk} that the wedge algebra of any classical $\mathcal{W}$-algebra can be obtained by simply dropping the non-linear terms in the Poisson brackets of the wedge modes \emph{without} taking the $\tilde{c}\to\infty$ limit. Therefore, one must have 
\begin{equation}
(\tilde{c}^{s''}_{ss'})_{\mathcal{W}_\infty^{cl}[\mu]}= h_{ss'}^{s''}\ ,
\label{eq:str_cst_DS}
\end{equation}
where $h_{ss'}^{s''}$ are the structure constants of $\wg{\mu}$ in the representation \eqref{eq:compact_comm_shss} for the commutators.
Using the classical version of eq.~\eqref{eq:corr1_mod} for $s=2$ and comparing with eqs.~\eqref{eq:hscom1} we get the desired relation
\begin{equation}\label{eq:2labels_mod}
\frac{\tilde{n}_2}{\tilde{c}} =  -\frac{(\mu -2) (\mu +1)}{8 (2 \mu-1)^2}\ .
\end{equation}
Thus, any classical $\mathcal{W}_\infty$-algebra parametrized by $\tilde{n}_2$ and $\tilde{c}$ can be realized as a DS reduction of  $\wg{\mu}$ provided $\tilde{n}_2$ is related to $\mu$ by eq.~\eqref{eq:2labels_mod}.
Reversing the logic, we also see that one can construct the DS reduction of $\wg{\mu}$ with the same techniques that we used to construct the most general classical $\mathcal{W}_\infty$-algebra.
For illustrative purposes we write down explicitly in app.~\ref{app:one_more} the first few Poisson brackets of $\mathcal{W}_\infty^{cl}[\mu]$.

Our observation that the most general classical and quantum $\mathcal{W}_\infty$-algebras have the same number of parameters implies that all classical $\mathcal{W}_\infty$-algebras can be recovered by taking the
classical limit of quantum $\mathcal{W}_\infty$-algebras. Let us briefly recall how to take the classical limit.
First, we need to go from OPEs to ``equal time'' commutators. If we want our ``space coordinate'' to live on the real line, then we must use the following prescription\footnote{We prefer to define equal time commutators on a real line rather then a circle in order to avoid performing a holomorphic transformation from
the plane to the cylinder under which $T$ picks up a shift of $-c/24$.
}
\begin{equation*}
\int_\mathbb{R} dx\, \epsilon(x)[W^{s\,\alpha}(x),W^{s'\,\alpha'}(y)] := \oint_y \frac{dz}{2\pi i}\,\epsilon(z)W^{s\,\alpha}(z) W^{s'\,\alpha'}(y)\ ,\quad y\in \mathbb{R}\ .
\end{equation*}
Using eq.~\eqref{eq:ope_not}, we get for the commutators
\begin{equation*}
[W^{s\,\alpha}(x),W^{s'\,\alpha'}(y)]=\sum_{n=1}^{s_1+\frac{\alpha_1}{2}+s_2+\frac{\alpha_2}{2}} [W^{s_1\, \alpha_1} W^{s_2\,\alpha_2}]_n(y) \frac{(-1)^{n-1}\partial_x^{n-1}\delta(x-y)}{(n-1)!}\ .
\end{equation*}
To take the classical limit we first rescale the generators and fundamental parameters of the $\mathcal{W}_\infty$-algebra by a power of $\hbar$
\begin{equation}\label{eq:cl_lim}
W^{s\,\alpha} = {\tilde{W}^{s\,\alpha}}\,{\hbar}^{-1}\ ,\quad n_2={\tilde{n}_2}\,{\hbar}^{-1}\ ,\quad c={\tilde{c}}\,{\hbar}^{-1}\ .
\end{equation}
If we now keep the rescaled generators $\tilde{W}^{s\,\alpha}$ fixed and expand their commutators in powers of $\hbar$
%
%
then the expansion starts at order $\mathcal{O}(\hbar)$.
This is a consequence of the Jacobi identities (\ref{jacobi_simple}, \ref{jacobi_higher}), which imply the following scaling for the structure constants
\begin{equation}\label{eq:scaling_str}
n_{s}\sim {\tilde{n}_{s}}\,{\hbar}^{-1}+\mathcal{O}(1)\ ,\quad c_{ss'}^{s''}\sim \tilde{c}_{ss'}^{s''}+ \mathcal{O}(\hbar),\quad a_{ss'}^{s''}\sim \tilde{a}_{ss'}^{s''}+\mathcal{O}(\hbar)\ .
\end{equation}
Thus,  in the limit $\hbar\to 0$ the commutator $[\tilde{W}^{s\,\alpha}(x),\tilde{W}^{s'\,\alpha'}(y)]$ vanishes  and we get
an algebra of functions with a Poisson bracket defined  by
\begin{equation}\label{eq:cl_pb}
\{\tilde{W}^{s\,\alpha}(x),\tilde{W}^{s'\,\alpha'}(y)\}:=\lim_{\hbar\to0}\frac{1}{\hbar}[\tilde{W}^{s\,\alpha}(x),\tilde{W}^{s'\,\alpha'}(y)]\ .
\end{equation}
It is useful to notice that the classical limit of any quantum $\mathcal{W}_\infty$-algebra whose parameter $n_2$ scales like~\eqref{eq:wedge_match} gives $\mathcal{W}^{cl}_\infty[\mu]$, i.e.\
$n_2/c$ reproduces \eqref{eq:2labels_mod} in the limit $\hbar\to0$.


\section{Minimal representations}\label{sec:min}


The analysis in~\cite{Creutzig:2012ar} of the 't~Hooft limit of the
partition functions of the minimal models~\eqref{eq:n1_cosets} suggests that the matter fields of the dual classical $\wg{\mu}$ higher spin bulk theory transform in specific representations
of the algebra of asymptotic symmetries $\mathcal{W}^{cl}_\infty[\mu]$.
These representations are called \emph{minimal} and they are defined by a Virasoro character of the form
\begin{equation}
 \label{eq:minrep}
\chi_{\text{min}}(q) 
=  q^h \, \frac{1+q^{\frac{1}{2}}}{1-q} \, \frac{\prod_{s\in\mathbb{N}+\frac{1}{2}}\prod_{n=0}^{\infty}(1+q^{n+s})}{\prod_{s\in2\mathbb{N}}\prod_{n=0}^\infty(1-q^{n+s})^2}
= q^h \, \frac{1+q^{\frac{1}{2}}}{1-q} \times \chi_{\infty} (q)\ .
\end{equation}
Their existence is a highly non-trivial fact, even classically, because infinitely many null vectors must appear at every level.
In this section we shall find convincing evidence for their existence, both at a classical and quantum level, and also compute the possible values of $h$.
This calculation can be seen as a holographic prediction for the quantum $1/c$ correction to the masses of the matter multiplets in the bulk theory.

In \eqref{eq:minrep}, we have factorized the character into two pieces --- 
the $\mathcal{W}_\infty$ vacuum character~\eqref{eq:ch_vac} and the character of an $\mathfrak{osp}(1|2)=\{L_0,L_{\pm1},G_{\pm 1/2}\}$  Verma module that lifts to a representation of the wedge algebra $\mathfrak{shs}^\sigma[\mu]$.
For the lift to be possible the 
$\mathfrak{osp}(1|2)$ Casimir must be $\mu(\mu-1)/4$, see app.~\ref{shssigma}.
The two Verma modules with this value of the Casimir have $L_0$ lowest weights $\mu/2$ and $(1-\mu)/2$.
Thus, the conformal dimension of the minimal representation must be either
\begin{equation}\label{hlimit}
h = \frac{\mu}{2} +\mathcal{O}(c^{-1}) \quad \text{or}\quad h = \frac{1-\mu}{2} +\mathcal{O}(c^{-1})
\end{equation}
if the $c\to\infty$ limit is well-defined. Quantum $1/c$ corrections can appear because the wedge algebra becomes an honest subalgebra only in the $c\to\infty$ limit.
We can now interpret the factorization~\eqref{eq:minrep} in the following way:
the minimal representation  is the ``extension'' beyond the wedge of an $\mathfrak{shs}^\sigma[\mu]$ representation lifted  from $\mathfrak{osp}(1|2)$;
under this extension the minimal representation inherits all the null vectors of $\mathfrak{shs}^\sigma[\mu]$ and their $\mathcal{W}_\infty$-descendants span the
null space of the minimal representation.\footnote{Actually, the analysis of~\cite{Creutzig:2012ar} suggests that the wedge algebra entirely determines the null vector structure
not only for the minimal representation, but also for all representations generated by the fusion product of a single minimal representation.}

We shall now exploit the rich null vector structure of the minimal representation and determine the possible values of $h$ at finite $c$
or, equivalently, $n_2$ as a function of $h$ and $c$ using the same approach as in \cite{Hornfeck:1993kp, Gaberdiel:2012, Candu:2012tr, Candu:2012ne}.
Decomposing the character \eqref{eq:minrep} in terms of $\mathcal{N}=1$ Verma module characters~\eqref{eq:n1_chs} of conformal dimensions $\tfrac{1}{2}\mathbb{N}+h$
\be
\chi_{\text{min}}(q) = \sum_{s\in\frac{1}{2}\mathbb{N}} d_{\text{min}}(s) \, \chi_{h+s} (q) 
= q^h \frac{1+q^{\frac{1}{2}}}{1-q}\chi_{0}(q) \sum_{s\in\frac{1}{2}\mathbb{N}} d_{\text{min}}(s) \, q^{s}\ , 
\ee
we get  the counting  function of $\mathcal{N}=1$ primaries in the minimal representation
\begin{equation}\label{eq:minpct}
D_{\mathrm{min}}(q) =  \sum_{s\in\frac{1}{2}\mathbb{N}} d_{\text{min}}(s) \, q^{s} = 
\frac{\chi_\infty(q)}{\chi_0(q)}= 1+q^2+ q^{\frac{5}{2}} + q^3 + 2\, q^{\frac{7}{2}}+\cdots\ .
\end{equation}
%
Let us denote the  $\mathcal{N}=1$ primary multiplet corresponding to the first term in the above expansion by  $P$.
According to the null vector discussion above,   the minimal representation is realized on the space of composite fields built out of normal ordered products between the $\mathcal{W}_\infty$-generators and their derivatives and a single $P$ factor or its derivatives.

We can now make the most general ansatz  for the OPEs between the first few $\mathcal{W}_\infty$-generators and $P$
\begin{align} 
W^{(2)} \times P &\sim w_2 P \ ,  \label{eq:minAnsatz1} \\
W^{(\frac{7}{2})} \times P &\sim w_{\frac{7}{2}} P + a_2 P^{(2)} + a_{\frac{5}{2}} P^{(\frac{5}{2})} + a_3 P^{(3)} \ ,\label{eq:minAnsatz2}\\
W^{(4)} \times P &\sim w_4 P + b_2 P^{(2)} + b_{\frac{5}{2}} P^{(\frac{5}{2})} + b_3 P^{(3)} + b_{\frac{7}{2},1} P^{(\frac{7}{2},1)}+ b_{\frac{7}{2},2} 
P^{(\frac{7}{2},2)} \ . \label{eq:minAnsatz3}
\end{align}
Here $P^{(s,i)}$ is the $i$-th  $\mathcal{N}=1$ primary supermultiplet of conformal dimension $h+s$ predicted by eq.~\eqref{eq:minpct}.
Their $\mathcal{N}=1$ primary components   are determined by
\begin{align*}
P^{2\, 0} & = W^{2\, 0} P^0 + \cdots \ , \\
P^{\frac{5}{2}\, 0} & = W^{2\, 0} P^1 - \frac{h}{2}\, W^{2\, 1} P^0 + \cdots \ ,\\
P^{3\, 0} & =  W^{2\, 1} P^1 - 4\, W^{2\, 0} P^0{}' + 2\, h\, W^{2\, 0}{}' P^0 + \cdots  \ ,\\
P^{\frac{7}{2},1\,0}& =W^{2\,0}{}'P^1+W^{2\,1}P^0{}'+
\frac{48 h (3+h)}{6+5 c-33 h+10 c h+30 h^2} G(W^{2\,0}P^{0})\,\\
&
\quad-\frac{2h \left(30+c+3 h+2 c h+6 h^2\right)}{6+5c-33 h+10 c h+30 h^2}W^{2\,1}{}'P^0 -
\frac{4 (6+5 c+27 h )}{6+5c-33 h+10 c h+30 h^2} W^{2\,0}P^1{}'+\cdots\ ,\\
P^{\frac{7}{2},2\,0}& = W^{\frac{7}{2}\,0} P+\cdots\ ,
\end{align*}
where we have only written down the dominant terms (see app.~\ref{sec:composite} for
an explanation of the term dominant). By solving the Jacobi identity
\be\label{ope22p}
[W^{(2)} \; ,\; W^{(2)} \; ,\; P] \ ,
\ee
we can then obtain the desired expression for $n_2$ in terms of $c$ and $h$
%
\be  \label{eq:n2}
n_2 = -\frac{c \left(2 c h-2 c-12 h^2+9 h\right) \left(2 c h+c+6 h^2-9 h\right) \left( c_{2 2}^2 \right)^2}{8 \left(4 c h-c-6 h^2\right)^2} \ ,
\ee
together with analogous expressions for the structure constants of the first two OPEs (\ref{eq:minAnsatz1}, \ref{eq:minAnsatz2})
\begin{align}\notag
w_2 &= \frac{h \left(2 c h+c+6 h^2-9 h\right) c_{2 2}^{2}}{2 \left(4 c h-c-6 h^2\right)} \ , \\ \notag
w_{\frac{7}{2}} & = 
-\frac{9 (2 c-3) (h+1) (c-9 h+3) (c+6 h) \left(2 c h+c+6 h^2-9 h\right) \left( c_{2 2}^2 \right)^2}{2 (4 c+21) (5 c+6) \left(4 c h-c-6 h^2\right)^2
 c_{2 2}^{\frac{7}{2}}  }
\ , \\ \notag
a_2 & = \frac{18 h (h+1) (c-9 h+3) c_{2 2}^2 }{(5 c+6) (h+2) \left(4 c h-c-6 h^2\right) c_{2 2}^{\frac{7}{2}}} \ , \\ \notag
a_{\frac{5}{2}} & = \frac{9 \left(10 c h^2+19 c h-6 c-24 h^2-6 h\right) c_{2 2}^2}{(5 c+6) (h+2) \left(4 c h-c-6 h^2\right) c_{2 2}^{\frac{7}{2}}} \ , \\
a_3 & = -\frac{27 (h+1) (2 h-1) (c+6 h) c_{2 2}^2}{(5 c+6) (2 h +5) \left(4 c h-c-6 h^2\right) c_{2 2}^{\frac{7}{2}}} \ .
\label{eq:min_str_const}
\end{align}
We have also checked that the Jacobi identity
\begin{equation*}
[W^{(2)} \; ,\; W^{(\frac{7}{2})} \; ,\; P]
\end{equation*}
fixes uniquely all the structure constants in the last OPE~\eqref{eq:minAnsatz3}.
In the following we shall make use only of  $w_4$, which we write down explicitly
\begin{align}\notag
w_4 &= h(2c-3)(2 c h+c+6 h^2-9 h)\left(c_{22}^2\right)^2 \Big[(2c+29)(4 c h-c-6 h^2)^2 c_{2\frac{7}{2}}^4\Big]^{-1}\Big\{
27 (h+1)\\ \notag
&\times (2 c h+3 c+46 h^2-35 h)
(c+3-9 h) (c+6 h) c_{22}^2 \Big[56  (5 c+6)
(4 c h-c-6 h^2) c_{22}^{\frac{7}{2}}\Big]^{-1}\\ \notag
&-(140 c^2 h^2+62 c^2 h-22 c^2+420
c h^3-144 c h^2+609 c h-129 c+8370 h^3-8343 h^2\\
&+1305 h) a_{2\frac{7}{2}}^4\Big[28 c(10 c-7)\Big]^{-1}
\Big\}\ .
\label{eq:w4}
\end{align}

We expect  this pattern to continue at higher levels and shall assume in the following that the Jacobi identities 
$[W^{(s_1)}\, ,\, W^{(s_2)}\, ,\, P]$ determine  the structure constants of all the OPEs $W^{(s)}\times P$ \emph{uniquely} in terms of  $h$ and $c$.
Under this assumption, the $\mathcal{W}_\infty$-algebra has exactly \emph{four} minimal representations, for generic values of $n_2$ and $c$, corresponding to the four solutions of $h$
in   eq.~\eqref{eq:n2}.
%

We can use eq.~\eqref{eq:wedge_match} to check that in the $c\to\infty$ limit  two of the minimal representations reproduce the conformal dimensions~\eqref{hlimit}
 predicted by the wedge algebra,
while the other two representations have conformal dimensions diverging as
\begin{equation}\label{eq:div_h}
h = \frac{c}{3(1-\mu)}+\mathcal{O}(c^{0})\ ,\qquad
h = \frac{c}{3\mu}+\mathcal{O}(c^{0})\ .
\end{equation}
%

One can repeat the above minimal representation analysis in the case of the classical algebra $\mathcal{W}^{cl}_\infty[\mu]$.
We have explicitly checked that the classical version of the Jacobi identity~\eqref{ope22p} reproduces the classical limits of eqs.~(\ref{eq:n2}, \ref{eq:min_str_const}, \ref{eq:w4}).
In particular, we get
\begin{equation*}
\tilde{n}_2 =-\frac{\tilde{c} ( \tilde{h}-1) (2  \tilde{h}+1)
\left( \tilde{c}_{2 2}^2 \right)^2}{4 (4 \tilde{h}-1)^2}\ .
\end{equation*}
Thus, under similar assumptions to the quantum case, the algebra $\mathcal{W}^{cl}_\infty[\mu]$ has only two minimal representations.
Comparing with~\eqref{eq:2labels_mod}, which requires normalizing  $\tilde{c}_{22}^2=1$, we get the expected  conformal dimensions
\begin{equation}\label{eq:hmin_cl}
\tilde{h} = \frac{\mu}{2} \ ,\qquad  \tilde{h} = \frac{1-\mu}{2}\ .
\end{equation}
The classical higher spin charges are
\begin{align}\notag
\tilde{w}_2 &= \frac{\tilde{h} (2 \tilde{h}+1) \tilde{c}_{2 2}^{2}}{2 (4 \tilde{h}-1)} \ , \\ \notag
\tilde{w}_{\frac{7}{2}} & = 
-\frac{9  (\tilde{h}+1) (2  \tilde{h}+1) \left( \tilde{c}_{2 2}^2 \right)^2}{20 (4  \tilde{h}-1)^2
 \tilde{c}_{2 2}^{\frac{7}{2}}  }
\ , \\ 
\tilde{w}_4 &= \frac{27\tilde{h}(\tilde{h}+1)(2\tilde{h}+1)(2\tilde{h}+3)\left(\tilde{c}_{22}^2\right)^2}{280(4\tilde{h}-1)^3\tilde{c}_{22}^{\frac{7}{2}}\tilde{c}_{2\frac{7}{2}}^4}\ .
\label{eq:cl_hs}
\end{align}

In conclusion, let us mention that eqs.~(\ref{eq:minAnsatz1}--\ref{eq:min_str_const})
hold for any $\mathcal{N}=1$ $\mathcal{W}$-algebra whose spin content begins with
\begin{align} \label{eq:firstSpins}
(\tfrac{3}{2}, 2)\ ,\ (2,\tfrac{5}{2})\ ,\ (\tfrac{7}{2},4)\ ,\ \ldots\ .
\end{align}
However, in this  case the precise definition of the minimal representation must be clarified and its existence justified independently.
We shall say that a representation of a \emph{general} $\mathcal{N}=1$ $\mathcal{W}$-algebra is minimal if it can be realized on the space of composite fields built out of any number of $\mathcal{W}$-algebra generators and their derivatives and a single superprimary multiplet or its derivatives.


\section{Truncations}
\label{sec:tr}

We have seen in sec.~\ref{sec:classical} that $\mathcal{W}_{\infty}^{cl}[\mu]$ is isomorphic to
the classical DS reduction of $\wg{\mu}$.
We shall use this isomorphism to study the truncation properties of $\mathcal{W}_{\infty}$ based on 
the truncation properties of $\wg{\mu}$.
As a preparation for the quantum case, it is useful to consider the classical case first.

\subsection{Classical analysis}
\label{sec:trunc_cl}


We show in app.~\ref{shssigma} that at positive integer values of $\mu=N$ the Lie superalgebra $\mathfrak{shs}^\sigma[\mu]\simeq \mathfrak{shs}^\sigma[1-\mu]$ acquires an ideal $\chi^\sigma_N$ such that
the quotient by this ideal can be identified with
\begin{equation}
\wg{N}/\chi_N^{\sigma} = \wg{1-N}/\chi_{1-N}^{\sigma}=\begin{cases}
\mathfrak{osp}(N \vert N-1)=B(n,n) & \text{for $N=2n+1$}\\
\mathfrak{osp}(N-1 \vert N)=B(n-1,n) & \text{for $N=2n$ .}
\end{cases}
\end{equation}
For this reason, see also eq.~\eqref{eq:str_cst_DS}, the  DS reduction should also truncate
%
%
\begin{equation}
\Winfcl{N}/\mathcal{I}_N =\Winfcl{1-N}/\mathcal{I}_{1-N} = \begin{cases} \mathcal{WB}^{cl}(n,n) &\text{for } N = 2n+1\\
\mathcal{WB}^{cl}(n-1,n) &\text{for } N = 2n \ ,
\end{cases} 
\label{eq:b_trunc}
\end{equation}
where $\mathcal{I}_N$ is the maximal ideal of $\Winfcl{N}$ and
we have denoted by  $\mathcal{WB}^{cl}(n,n)$ and $\mathcal{WB}^{cl}(n-1,n)$ the classical DS reductions of the Lie superalgebras 
$B(n,n)$ and $B(n-1,n)$ w.r.t.\ the \emph{principal} embedding of $\mathfrak{osp}(1|2)$.\footnote{Only the superalgebras $A(n+1,n),\ B(n,n),\ B(n-1,n),\ D(n, n), \ D(n+1, n)$ and $ D(2,1;\alpha)$ admit such a principal embedding, see \cite{Delduc:1991sg, Evans:1990qq}.
}
The spectrum of their generators 
\begin{align}
\mathcal{WB}^{cl}(n,n) &: \ (\tfrac{3}{2},2) \ , \ (2,\tfrac{5}{2})  \  , \ (\tfrac{7}{2},4) \  , \ \dots  \  , \ (2n, 2n + \tfrac{1}{2}) \ , \notag \\
\mathcal{WB}^{cl}(n-1,n) &: \ (\tfrac{3}{2},2)  \  , \ (2,\tfrac{5}{2})  \  , \ (\tfrac{7}{2},4) \  , \ \dots  \  , \ (2n-\tfrac{1}{2},2n)
\label{eq:wb_spec}
\end{align}
has been computed in \cite{Delduc:1991sg, Evans:1990qq} and is, of course, consistent with the fact that these algebras are truncations of $\mathcal{W}^{cl}_{\infty}[\mu]$.

Formulas for the central charge $\tilde{c}$ and the spectrum of conformal dimensions $\tilde{h}$ of the DS reductions~\eqref{eq:wb_spec} 
have been derived in \cite{Evans:1990qq} from a Toda theory point of view
\be \label{eq:classicalDS}
\tilde{c} = - 12\tilde{\alpha}_-^2 \left(\frac{\rho^{\vee}}{2}, \frac{\rho^{\vee}}{2} \right) \ ,\qquad 
\tilde{h}(\Lambda_+) = -  \left(\Lambda_+, \frac{\rho}{2}^{\vee}\right) \ ,
\ee
where $\tilde{\alpha}_-$ parametrizes the level of the DS reduction, while 
$\Lambda_+$ is a weight of the respective Lie superalgebra and $(\cdot\,,\cdot)$ is the scalar product in weight space.
Recall that Lie superalgebras  (usually)  have many inequivalent simple root systems. However, the principal 
embedding of $\mathfrak{osp}(1|2)$ singles out a special simple root system $\{\alpha_i\}_{i=1}^r$, consisting purely of fermionic roots. 
W.r.t.\ this simple root system the Weyl covector is then defined  by the property
\begin{equation}
(\rho^\vee,\alpha_i) = 1\ ,
\label{eq:def_wc}
\end{equation}
see app.~\ref{sec:lsa} for more details.
We note that eq.~\eqref{eq:classicalDS} holds for all DS reductions w.r.t.\ a principal $\mathfrak{osp}(1|2)$ embedding and not only for the algebras $\mathcal{WB}^{cl}(n,n)$ and $\mathcal{WB}^{cl}(n-1,n)$.

The spectrum~\eqref{eq:classicalDS} is strictly speaking continuous.
A discrete spectrum will arise  only after restricting to $\mathcal{W}$-algebra representations with sufficiently many null vectors, i.e.\ fully degenerate representations.
Although a classification of fully degenerate representations for the classical DS reductions w.r.t.\ a principal $\mathfrak{osp}(1|2)$ embedding is not yet available, it is tempting to believe that, just as in the bosonic case,  these will be parametrized by dominant\footnote{A weight of a Lie (super)algebra is said to be dominant if it is the highest weight of a finite dimensional representation. See app.~\ref{sec:lsa} for more details on the dominance condition.} weights $\Lambda_+$, see \cite{Evans:1990qq}.
Put differently, it is natural to expect that fully degenerate representations are, in a sense, extensions beyond the wedge of finite dimensional representations of the wedge algebra.
The following claim relies on this technical assumption.

Let us identify the DS reductions \eqref{eq:wb_spec}  with a truncation $\Winfcl{N}/\mathcal{I}_N$ at a positive integer value of $N$ by the isomorphism~\eqref{eq:b_trunc}.
Then we claim that the irreducible DS-representation labelled by the highest weight $\Lambda_+=v$ of the vector representation (see app.~\ref{sec:lsa})
must be identified with the representation of $\Winfcl{N}/\mathcal{I}_N$ defined by the quotient
\begin{equation}
M_-/ (\mathcal{I}_N \cdot M_-)\ ,
\label{eq:trunc_min}
\end{equation}
 where $M_-$ is the minimal representation of $\Winfcl{N}$
of lowest conformal dimension $(1-N)/2$.
Indeed, inserting $\Lambda_+=v$ in eq.~\eqref{eq:classicalDS} and comparing with eq.~\eqref{eq:hmin_cl} we see that the conformal dimensions of the ground states agree
\begin{equation}
\tilde{h}(v) = \frac{1-N}{2}\ ,\qquad N = \begin{cases} 2n+1 & \text{for } \mathcal{WB}^{cl}(n,n)\\ 2n & \text{for } \mathcal{WB}^{cl}(n-1,n) \ . \end{cases}\ 
\label{eq:cl_1st_min}
\end{equation}
Moreover, we prove in app.~\ref{shssigma} that the wedge algebra of  $\Winfcl{N}/\mathcal{I}_N$, which is $\mathfrak{osp}(N|N-1)$ for $N$ odd and $\mathfrak{osp}(N-1|N)$ for $N$ even, generates the vector
representation when acting on the ground state of~\eqref{eq:trunc_min}.

What about the minimal representation of $\Winfcl{N}$ of conformal dimension $\tilde{h}=N/2$?
In fact, this representation, which we will denote by $M_+$, cannot truncate to a non-trivial representation of $\Winfcl{N}/\mathcal{I}_N$
because the ideal $\mathcal{I}_N$ does not act trivially on its ground state.\footnote{The wedge algebra analysis in app.~\ref{shssigma} suggests that $M_+=\mathcal{I}_N\cdot M_-$.}
In the cases $N=2,3,4$ we have checked explicitly  with the help of eqs.~\eqref{eq:cl_hs} that the  higher spin charges of the generators $\tilde{W}^{(s)}\in\mathcal{I}_N$ with  $s\leq 4$ do not vanish
for  $M_+$, contrary to $M_-$. This is a sign that the quotient $M_+/(\mathcal{I}_N\cdot M_+)$ is trivial, while the quotient $M_-/(\mathcal{I}_N\cdot M_-)$
is not.

One can show, however,  that there is a \emph{different} truncation of $\Winfcl{N}$ such that the minimal representation $M_+$ does survive!
It can be described in terms of the  DS reductions of the Lie superalgebras $D(n,n)$ and $D(n+1,n)$, which we denote by
\begin{align}\label{eq:d_DS}
\mathcal{WD}^{cl}(n,n) &: \ (\tfrac{3}{2},2) \ , \ (2,\tfrac{5}{2}) \  , \ (\tfrac{7}{2},4)\  , \ \dots \  , \ (2n-\tfrac{1}{2},2n)\  , \ (n,n + \tfrac{1}{2}) \ , \\ \notag
\mathcal{WD}^{cl}(n+1,n) &: \ (\tfrac{3}{2},2) \  , \ (2,\tfrac{5}{2}) \  , \ (\tfrac{7}{2},4)\  , \ \dots \  , \ (2n,2n+\tfrac{1}{2}) \  , \ (n+\tfrac{1}{2},n+1) \ , 
\end{align}
see \cite{Delduc:1991sg, Evans:1990qq}.
More precisely, the truncations are realized as  orbifolds of the above DS reductions by the
sign flip automorphism of the last generator in~\eqref{eq:d_DS}.
This automorphism, which we denote by $\tau$, is inherited from the outer automorphism of the $D$-type Lie superalgebras.

To understand our claim, first notice that the orbifold subalgebras are generated by the subset of $\tau$-even generators in~\eqref{eq:d_DS}
plus the bilinears in the last $\tau$-odd generator and its derivatives
which are not themselves total derivatives.
An easy counting reveals that this generating spectrum is of the form~\eqref{eq:Winf_spec}, hence   $\mathcal{WD}^{cl}(n,n)^\tau$ and $\mathcal{WD}^{cl}(n+1,n)^\tau$ must be quotients of
$\Winfcl{\mu}$.\footnote{The counting argument for a similar algebra is spelled out in sec.~\ref{sec:counting}.}

To determine the value of $\mu$ for which the truncation takes place, notice that the odd generator in \eqref{eq:d_DS} generates, in the sense of the definition given at the end of sec.~\ref{sec:min}, a minimal representation of the orbifold algebra
of conformal dimension
\begin{equation}
\tilde{h} = \frac{N}{2}\ ,\qquad N = \begin{cases} 2n & \text{for } \mathcal{WD}^{cl}(n,n) \\ 2n+1 & \text{for } \mathcal{WD}^{cl}(n+1,n) \ . \end{cases}
\label{eq:cld_1st_min}
\end{equation}
The minimality of this representation follows from the fact that the Poisson bracket between an even generator  and the odd generator must be, for dimensional reasons, linear in the latter,
 which is equivalent to the defining property of the minimal representation, see sec.~\ref{sec:min}.
Comparing to eqs.~\eqref{eq:hmin_cl} we see that for integer values of $\mu$ there must be an ideal $\mathcal{J}_\mu\neq \mathcal{I}_\mu$  such that
the following truncations occur
\begin{equation}
\Winfcl{N}/\mathcal{J}_N =\Winfcl{1-N}/\mathcal{J}_{1-N} = \begin{cases} 
\mathcal{WD}^{cl}(n,n)^\tau &\text{for } N = 2n \\ \mathcal{WD}^{cl}(n+1,n)^\tau &\text{for } N = 2n+1 \ .
\end{cases} 
\label{eq:d_trunc}
\end{equation}
This differs from~\eqref{eq:b_trunc} in  that  the minimal representation $M_+$ survives the truncation.

In conclusion, we  summarize the truncation properties of the algebra $\Winfcl{\mu}$  by the following diagram
\begin{equation}
\begin{tikzpicture}[scale=.9]
\node (W) at (0,0) {$\mathcal{W}^{cl}_{\infty}[\mu]$};
\node at (3.5,0) {$\mu=2n+1\,, -2n$};
\node at (-3.5,0) {$\mu=2n\,, 1-2n$};
\draw[->>,>=stealth, label="$\mu=2n$"] (W.north east)--  (3,1.5) node [above right]{$\mathcal{WB}^{cl}(n,n)$};
\draw[->>,>=stealth] (W.south east)--  (3,-1.5) node [below right]{$\mathcal{WD}^{cl}(n+1,n)^\tau$};
\draw[->>,>=stealth] (W.north west)--   (-3,1.5) node [above left]{$\mathcal{WB}^{cl}(n-1,n)$};
\draw[->>,>=stealth] (W.south west)--  (-3,-1.5) node [below left]{$\mathcal{WD}^{cl}(n,n)^\tau$};
\end{tikzpicture}
\label{eq:trunc_cl}
\end{equation}
where  the double headed arrows denote projection homomorphisms.\footnote{This is consistent with the fact that $D(n+1,n)^\tau\simeq B(n,n)$ and $D(n,n)^\tau\simeq B(n-1,n)$.}


\subsection{The algebra \texorpdfstring{$\Winf{\mu}$}{Winfty[mu]} and its truncations}
\label{sec:winfmu}


The truncations~\eqref{eq:trunc_cl} should generalize to the quantum case.
The goal of this section is to define an algebra $\Winf{\mu}$ that parametrizes the space of $\mathcal{W}_\infty$-algebras and has the property  that (i) it truncates to (orbifolds of) the quantum DS reductions
\begin{equation}
\mathcal{WB}(n,n)\ ,\quad \mathcal{WB}(n-1,n)\ ,\quad \mathcal{WD}(n,n)\ ,\quad \mathcal{WD}(n+1,n)
\label{eq:qds}
\end{equation}
at simple values of $\mu$ (e.g.\ integer when possible) and (ii) it reproduces $\Winfcl{\mu}$ in the classical limit.
These are natural properties that one would expect from a generalized quantum DS reduction of $\wg{\mu}$.

The central charges and spectra of the quantum DS reductions~\eqref{eq:qds} were given in \cite{Evans:1990qq}
\be\label{QDS_cc_spec}
c = \frac{3}{2} r - 12 \left( \alpha_-\,{\frac{\rho^{\vee}}{2}}  +  \alpha_{+}\,\rho\right)^2 \ ,\qquad
h(\Lambda) = \frac{1}{2} \left(\Lambda,\Lambda +  \alpha_-\,\rho^{\vee} + 2\, \alpha_{+}\, \rho \right) \ ,
\ee
where $\alpha_-$ parametrizes the level of the DS reduction and $\alpha_+:=-1/\alpha_-$, while $\Lambda$ is a weight  of the corresponding Lie superalgebra.
Explicit expressions for the Weyl vector  $\rho$ and the Weyl covector $\rho^\vee$, which are
defined w.r.t.\ a purely fermionic simple root system $\{\alpha_i\}_{i=1}^r$  by the properties
\begin{equation}
(\rho,\alpha_i) = \frac{(\alpha_i,\alpha_i)}{2}
\label{eq:def_wv}
\end{equation}
and \eqref{eq:def_wc}, are given in app.~\ref{sec:lsa}.
The classical limit~\eqref{eq:classicalDS} is obtained by setting $\Lambda = \alpha_+\Lambda_+$ and then by taking the limit
$\alpha_+,\hbar\to 0$ with $\tilde{\alpha}^2_-=\hbar\alpha^2_-$ kept fixed.

We have seen in sec.~\ref{sec:trunc_cl} that for the algebras $\mathcal{WB}(n,n)$ and $\mathcal{WB}(n-1,n)$ 
the representation parametrized by $\Lambda = \alpha_+ v$ is minimal, where  $v$ is the highest weight of the vector representation.
For the  orbifold algebras $\mathcal{WD}(n,n)^\tau$ and $\mathcal{WD}(n+1,n)^\tau$  there is no reason to assume that $\Lambda = \alpha_+ v$ corresponds to a minimal representation.
A minimal representation is provided in this case by the last current in~\eqref{eq:d_DS}.
Now, in the quantum case, the conformal dimensions of these minimal representations are given by 
\begin{equation}\label{eq:expl_h_qds}
h = \frac{1-\mu}{2}\ ,\qquad  \text{where}\quad \begin{cases}
\mu = 2n+1 & \text{for } \mathcal{WB}(n,n)\\
\mu = 2n-2\alpha_+^2 & \text{for } \mathcal{WB}(n-1,n)\\
\mu = 1-2n & \text{for } \mathcal{WD}(n,n)\\
\mu = -2n & \text{for } \mathcal{WD}(n+1,n) \ ,
\end{cases}
\end{equation}
where we have used eq.~\eqref{QDS_cc_spec} in the first two cases.
Remarkably, we can also write the central charges of the above algebras in a single formula
\begin{equation}\label{eq:expl_c_qds}
c = \frac{3}{2}(1-\mu)(1+ \alpha_-^2\mu)\ .
\end{equation}

Hence, by plugging \eqref{eq:expl_h_qds} and \eqref{eq:expl_c_qds} into eq.~\eqref{eq:n2}, we obtain an expression for $n_2$ 
in terms of $\mu$ and $\alpha^2_+$ 
\begin{equation}
\frac{n_2}{c} = -\frac{(\mu-2+2\alpha_+^2)(\mu+1-\alpha_+^2)}{8(2\mu-1+\alpha_+^2)^2 } \left(c_{22}^2\right)^2\ .
\label{eq:def_n2_am}
\end{equation}
This enables us to define $\Winf{\mu}$ as the algebra  $\mathcal{W}^{n_2,c}_\infty$  with $c$ and $n_2$ given by eqs.~\eqref{eq:expl_c_qds} and \eqref{eq:def_n2_am}.
Comparing eq.~\eqref{eq:def_n2_am} to its classical version~\eqref{eq:2labels_mod} we see that the classical limit of $\Winf{\mu}$ coincides with $\Winfcl{\mu}$.
Moreover, it follows from eqs.~\eqref{eq:expl_c_qds} and \eqref{eq:def_n2_am} that the quantum version of the truncation diagram~\eqref{eq:trunc_cl} is given by
\begin{equation}
\begin{tikzpicture}[scale=.9]
\node (W) at (0,0) {$\mathcal{W}_{\infty}[\mu]$};
\draw[->>,>=stealth,label="$\mu=2n$"] (W.north east)-- node [sloped,below]{$\scriptstyle{\mu=2n+1}$} (3,1.5) node [above right]{$\mathcal{WB}(n,n)$};
\draw[->>,>=stealth] (W.south east)-- node [sloped,above]{$\scriptstyle{\mu=-2n}$} (3,-1.5) node [below right]{$\mathcal{WD}(n+1,n)^\tau\ .$};
\draw[->>,>=stealth] (W.north west)-- node [sloped,below]{$\scriptstyle{\mu=2n-2\alpha_+^2}$}   (-3,1.5) node [above left]{$\mathcal{WB}(n-1,n)$};
\draw[->>,>=stealth] (W.south west)-- node [sloped,above]{$\scriptstyle{\mu=1-2n}$} (-3,-1.5) node [below left]{$\mathcal{WD}(n,n)^\tau $};
\end{tikzpicture}
\label{eq:tr_q}
\end{equation}
Notice that only the value of $\mu$ corresponding to the $\mathcal{WB}(n-1,n)$ truncation receives a quantum correction.
Let us also mention that we have explicitly verified with the help of the Jacobi identities \eqref{eq:jacobi} that
$\Winf{\mu}$ truncates to $\mathcal{WB}(0,1)$, $\mathcal{WB}(1,1)$ and $\mathcal{WB}(1,2)$  at the expected values of $\mu$.

One can now use eq.~\eqref{eq:n2} again to  extract the conformal dimensions of all four minimal representations of $\Winf{\mu}$
\begin{equation}
h_1 = \frac{1-\mu}{2}\ ,\qquad h_2 = \frac{1-\mu}{2\alpha_+^2}\ ,\qquad h_3 = \frac{ \mu+\alpha_+^2}{2\alpha_+^2}\ ,\qquad h_4 = \frac{\mu+\alpha_+^2}{2}\ .
\label{eq:4min_winfm}
\end{equation}
Eliminating  $\alpha_-^2$ in favor of $c$ with the help of eq.~\eqref{eq:expl_c_qds},  we can rewrite the above conformal dimensions in terms of $\mu$ and $c$ as follows
\begin{equation} 
h_1 = \frac{1-\mu}{2}\ ,\qquad h_2 = \frac{2c+3\mu-3}{6\mu}\ ,\qquad h_3 = \frac{c}{3(1-\mu)}\ ,\qquad h_4 = \frac{c\mu}{2c+3\mu-3}\ .
\label{eq:4min_winfm_mc}
\end{equation}
They have the expected $c\to\infty$ behavior, see eqs.~(\ref{hlimit}, \ref{eq:div_h}).

For the truncations of $\mathcal{W}_\infty[\mu]$ to $\mathcal{WB}(0,1)$, $\mathcal{WB}(1,1)$ and $\mathcal{WB}(1,2)$  the higher spin charges~(\ref{eq:min_str_const}, \ref{eq:w4}) of the generators of spin $s\leq 4$ that belong to the ideal of
$\mathcal{W}_\infty[\mu]$ vanish only in the first three representations.
Thus, only the first three minimal representations survive the truncation to a $B$-type DS reduction.
The fact the last representation does not survive the truncation is already clear form the classical analysis of sec.~\ref{sec:trunc_cl}.
We shall see in sec.~\ref{sec:cosets} that this is also true for the $D$-type truncations.

In conclusion, let us mention that the parametrization of $\mathcal{W}_\infty$ in terms of the pair $(\mu,\alpha^2_+)$ is not unique.
In fact, there are generically four distinct pairs 
\begin{align}
(\mu,\alpha^2_+)\ ,\quad (1-\mu-\alpha_+^2, \alpha^2_+)\  ,\quad (-\alpha_-^2 \mu,\alpha_-^2)\ ,\quad (1+\alpha_-^2\mu-\alpha_-^2,\alpha_-^2)
\label{eq:self_d}
\end{align}
corresponding to the same  $\mathcal{W}_\infty$ algebra, i.e.\ the same values of $c$ and $n_2$.
Therefore, there are three other values of $\mu$ at which the truncations  eq.~\eqref{eq:tr_q} occur.
From the point of view of the $\Winf{\mu}$ algebra, the relations~\eqref{eq:self_d} give rise to self-dualities.
In particular, notice that the classical symmetry $\Winfcl{\mu}\simeq\Winfcl{1-\mu}$ acquires a quantum correction.


\section{\texorpdfstring{$\mathcal{N}=1$}{N=1} cosets}\label{sec:cosets}



It has been conjectured in \cite{Creutzig:2012ar}
that a certain $\mathcal{N}=1$ truncation of the $\mathcal{N}=2$ higher spin supergravity theory on AdS$_3$ of Prokushkin and Vasiliev \cite{Prokushkin:1998bq}
is dual to  the coset\footnote{Here $n$ and $k$ are positive integers and $\mathfrak{so}(2n)_1$ must be understood as the chiral algebra of $2n$  Neveu-Schwarz free fermions.}
\be \label{c1}
\frac{\mathfrak{so}(2n+1)_k \oplus\mathfrak{so}(2n)_1}{\mathfrak{so}(2n)_{k+1}} \ 
\ee
in the 't~Hooft like limit
\begin{equation}
n,k \rightarrow \infty \qquad\text{with}\qquad  \lambda := \frac{2n}{2n+k-1} \qquad \text{held fixed.}
\label{eq:thooft}
\end{equation}
The supergravity theory is essentially an $\wg{\lambda}$ Chern-Simons gauge theory  coupled to two real $\mathcal{N}=1$ matter multiplets with masses determined by $\lambda$.
Its  asymptotic symmetry algebra 
 can be identified with the DS reduction of $\wg{\lambda}$, i.e.\ with $\mathcal{W}^{cl}_\infty[\lambda]$, by the usual arguments \cite{Hanaki:2012yf, Henneaux:2012ny} (see also \cite{Henneaux:2010xg, Campoleoni:2010zq, Gaberdiel:2011wb, Campoleoni:2011hg} in the non-supersymmetric case).
The duality was put on solid grounds by matching the 1-loop partition functions on both sides,\footnote{Modulo  what is believed to be null  vectors appearing in the 't~Hooft limit of the CFT.}
which also provided indirect evidence for the agreement of symmetry algebras.

In this section we shall identify the chiral algebra of the coset~\eqref{c1} with the DS reduction of $\mathfrak{osp}(2n|2n)$, i.e.
\begin{equation}
\frac{\mathfrak{so}(2n+1)_k \oplus\mathfrak{so}(2n)_1}{\mathfrak{so}(2n)_{k+1}} \simeq \mathcal{WD}(n,n)\ .
\label{eq:ii1}
\end{equation}
This fact together with the truncation properties of the algebra $\Winf{\mu}$ derived in sec.~\ref{sec:winfmu} will allow us to formulate the quantum version of the holographic duality at finite $n$ and $k$.
In particular, the matching of symmetries  in the 't~Hooft limit  will become fully transparent. 

There are two other $\mathcal{N}=1$ cosets for which we can find a DS reformulation
\begin{align}\label{eq:i2i3}
\frac{\mathfrak{so}(2n+2)_k \oplus \mathfrak{so}(2n+1)_1}{\mathfrak{so}(2n+1)_{k+1}} &\cong \mathcal{WD}(n+1,n) \ ,\\[5pt]
\frac{\mathfrak{osp}(1\vert 2n)_k \oplus \mathfrak{sp}(2n)_{-1}}{\mathfrak{sp}(2n)_{k-1}} &\cong \mathcal{WB}(n,n) \notag
\end{align}
and generalize the holography statement.


\subsection{Coset higher spin currents}
\label{sec:counting}


Our  first step towards proving (\ref{eq:ii1}, \ref{eq:i2i3}) is to match the spin content of the generators of the two algebras.
Let us recall the counting \cite{Creutzig:2012ar}  of  the higher spin currents generating the chiral algebra of the coset \eqref{c1}.
Denote the  $\mathfrak{so}(2n+1)_k$ currents by $J^{ab}(z)=-J^{ba}(z)$, where $a,b=0,\dots,2n$, and the $\mathfrak{so}(2n)_1$ Majorana fermions  by $\psi^i(z)$, where $i=1,\dots,2n$.
They satisfy the OPEs
\begin{align*}
\psi^i(z)\psi^j(w)&\sim \frac{\delta^{ij}}{z-w}\ ,\\
J^{ab}(z) J^{cd}(w)&\sim \frac{k\delta^{bc}\delta^{ad}}{(z-w)^2}+\frac{\delta^{bc}J^{ad}(w)-\delta^{da}J^{cb}(w)}{z-w}-(c\leftrightarrow d) \ .
\end{align*}
The currents  $K^{ij}(z)=J^{ij}(z)+\psi^i\psi^{j}(z)$, where $i,j=1,\dots,2n$ and $i\neq j$, generate the $\mathfrak{so}(2n)_{k+1}$ current algebra in the denominator in~\eqref{c1}.
The coset algebra  is then defined as the space of normal ordered polynomials in $J^i:=J^{0i}$, $\psi^i$, $K^{ij}$ and their derivatives that are regular w.r.t.\ numerator currents $K^{ij}$.
Our definition  for the  normal ordering is
\begin{equation}
(AB)(w) = \oint_w \frac{dz}{2\pi i } \frac{A(z)B(w)}{z-w}\ .
\label{eq:no}
\end{equation}

Let us denote by $\tau$ the $\mathbb{Z}_2$ automorphism of \eqref{c1} inherited from the outer automorphism of $\mathfrak{so}(2n)$.
The latter is explicitly given by 
$$
\tau(J^i) = J^{\tau(i)}\ ,\quad \tau(\psi^i) = \psi^{\tau(i)}\ ,\quad \tau(K^{ij}) = K^{\tau(i)\tau(j)}\ ,
$$
where $\tau(i) = i$ for $i = 1,\dots,2n-2$ and
$$
\tau(2n-1)=2n\ ,\quad \tau(2n)=2n-1\ .
$$
Alternatively, we can view $\tau$ as an improper $\mathrm{O}(2n)$ transformation.
To better organize the counting of coset generators we shall  also consider its $\tau$-invariant subalgebra or $\tau$-orbifold
\begin{equation}
 \label{orb}
\left(\frac{\mathfrak{so}(2n+1)_k \oplus\mathfrak{so}(2n)_1}{\mathfrak{so}(2n)_{k+1}}\right)^\tau \ .
\end{equation}

The generators are most easily computed in the $k\to\infty$ limit,
where the coset algebra simplifies to the space of $\mathfrak{so}(2n)$ invariant normal ordered polynomials in the abelian currents $\tilde{J}^i=J^{i}/\sqrt{k}$, the free fermions $\psi^i$
and their derivatives~\cite{Bouwknegt:1992wg}.\footnote{Taking the limit does not affect the counting, because $k$ can be reintroduced at any moment by simply replacing the derivatives of $\tilde{J}^i$ and $\psi^i$ with $K^{ij}$-dependent  covariant derivatives,
see~\cite{deBoer:1993gd, Creutzig:2012ar}.}
Notice that $\tilde{J}^i$ and $\psi^i$ transform in the fundamental representation of $\mathfrak{so}(2n)$.
Classical invariant theory tells us that all $\mathrm{O}(2n)$ invariant polynomials are generated by the elementary invariants \cite{Weyl}
\begin{equation}\label{eq:inv_gen}
(\partial^m \tilde{J}^i\partial^n \tilde{J}^i)\ ,\quad (\partial^m\psi^i\partial^n \psi^i)\ ,\quad \partial^m\psi^i\partial^n \tilde{J}^i\ .
\end{equation}
Removing total derivatives, we can reduce the set of generators to 
\begin{equation}\label{eq:coset_currents}
(\tilde{J}^i\partial^{2m}\tilde{J}^i)\ ,\quad (\psi^i\partial^{2m+1}\psi^i)\ ,\quad \psi^i \partial^m \tilde{J}^i\ ,\qquad m=0,1,2,\dots\ .
\end{equation}
Among these, the  currents
\begin{equation}\label{eq:vir_free}
G\propto \psi^i \tilde{J}^i\ ,\qquad 
T = \frac{1}{2}(\tilde{J}^i\tilde{J}^i) - \frac{1}{2}(\psi^i\partial\psi^i)
\end{equation}
generate an $\mathcal{N}=1$ Virasoro subalgebra. With respect to the latter, the higher spin  currents~\eqref{eq:coset_currents} organize into supermultiplets according to the list of generators~\eqref{eq:Winf_spec} of the $\mathcal{W}_\infty$-algebra.
Thus, the coset orbifold~\eqref{orb} must be a quotient of $\mathcal{W}_\infty$.

Let us call the $\tau$-invariant currents~\eqref{eq:inv_gen} $\tau$-even or simply even. Besides~\eqref{eq:coset_currents}, the coset~\eqref{c1} also contains  currents that are $\mathrm{SO}(2n)$-invariant, but not $\mathrm{O}(2n)$ invariant.
We call them $\tau$-odd or simply odd because they change sign under the action of the improper $\mathrm{O}(2n)$ transformation~$\tau$.
Again, classical invariant theory tells us that all elementary odd invariants are of the form \cite{Weyl}
\begin{equation}
\sum_{\sigma}\varepsilon(\sigma)\partial^{m_1}\psi^{\sigma(1)}\cdots \partial^{m_l}\psi^{\sigma(l)}\partial^{m_{l+1}}\tilde{J}^{\sigma(l+1)}\cdots \partial^{m_{2n}}\tilde{J}^{\sigma(2n)}\ ,
\label{eq:pseudo_gen_all}
\end{equation}
where the sum is over all permutations $\sigma$  and $\varepsilon(\sigma)$ is the signature of $\sigma$.
All other odd invariants can be obtained by multiplying~\eqref{eq:pseudo_gen_all} with even invariants and taking linear combinations.
In fact, we have a stronger claim: all odd invariants can be represented as linear combinations of normal ordered products between even invariants and the derivatives of only two basic odd invariants 
\be\label{eq:pseudo_gen}
V^0 := \psi^{1} \cdots \psi^{2n}\ ,\qquad  V^1 := \sum_{i=1}^{2n} (-1)^{i-1}\psi^1\cdots \tilde{J}^i\cdots \psi^{2n}\ ,
\ee
which group together into  a supermultiplet $V$ of spin $(n,n + \tfrac{1}{2})$ w.r.t.\ the $\mathcal{N}=1$ Virasoro algebra \eqref{eq:vir_free}.
The proof of this claim is rather cumbersome, although straightforward, so we shall just present a few examples to illustrate the mechanism by which all elementary  odd invariants \eqref{eq:pseudo_gen_all} can be generated from $V^0$ and $V^1$
\begin{align}\label{eq:observ}
\text{spin } n+1:&& \partial V^0 &= \sum_i \psi^1 \cdots \partial \psi^i \cdots \psi^{2n}\ ,\\ \notag
\text{spin } n+\tfrac{3}{2}:
&& 
\partial V^1 & = \sum_i (-1)^{i-1}\psi^1 \cdots \partial \tilde{J}^i \cdots \psi^{2n}- \sum_{i\neq j} (-1)^{j}\psi^1 \cdots \partial\psi^{i}\cdots  \tilde{J}^j \cdots \psi^{2n}\ ,\\ \notag
&&((\psi^i \tilde{J}^i) V^0)&=  \sum_i (-1)^{i-1}\psi^1 \cdots \partial \tilde{J}^i \cdots \psi^{2n}\ ,\\ \notag
\text{spin } n+2:&& \partial^2 V^0 &=  \sum_i \psi^1 \cdots \partial^2\psi^i \cdots \psi^{2n}+2\sum_{i< j}\psi^{1}\cdots \partial\psi^i \cdots \partial\psi^j \cdots \psi^{2n}\ ,\\ \notag
&& ((\psi^i\partial \psi^i) V^0) &=  -\tfrac{3}{2}\sum_i \psi^1 \cdots \partial^2\psi^i \cdots \psi^{2n}\ ,\\ \notag
&& ((\psi^i \tilde{J}^i) V^1)&= \sum_{i< j}(-1)^{i+j}(\psi^1 \cdots \partial \tilde{J}^i \cdots  \tilde{J}^j\cdots \psi^{2n}-\psi^1 \cdots \tilde{J}^j \cdots  \partial \tilde{J}^i\cdots \psi^{2n})\\ \notag
&& &+ \tfrac{1}{2}\sum_{i} \psi^1 \cdots \partial^2\psi^1\cdots \psi^{2n}\ .
\end{align}
Thus, for the moment it appears  that the coset algebra \eqref{c1} is generated by the even currents \eqref{eq:inv_gen} and the odd currents~\eqref{eq:pseudo_gen}.
However, we can trade the currents~\eqref{eq:inv_gen} of spin $s\geq 2n$ for the following bilinears in $V$\footnote{Here and in eq.~\eqref{eq:observ} it is  important that we use the normal ordering~\eqref{eq:no}. Notice that for composite fields the latter differs from the standard free field normal ordering.}
\begin{equation}
(V^0 \partial^{2m}V^0)\ ,\quad (V^1 \partial^{2m+1}V^1)\ ,\quad (V^0 \partial^mV^1)\ , \quad m=0,1,2,\dots\ ,
\label{eq:bilnn}
\end{equation}
where the shift in the order of derivatives of the first two terms is due to the opposite statistics of $V^0$ and $V^1$.
Hence, the coset algebra~\eqref{c1} must be an $\mathcal{N}=1$ $\mathcal{W}$-algebra generated by supermultiplets of spins
\begin{equation}\label{eq:spec_dnn}
(\tfrac{3}{2},2) \ , \ (2,\tfrac{5}{2}) \  , \ (\tfrac{7}{2},4)\  , \ \dots \  , \ (2n-\tfrac{1}{2},2n)\  , \ (n,n + \tfrac{1}{2}) \ .
\end{equation}
%


\subsection{Cosets as DS reductions}\label{sec:coset-ds}




Notice that \eqref{eq:spec_dnn} is precisely the spectrum of generators of the algebra $\mathcal{WD}(n,n)$,  see~\eqref{eq:d_DS}.
Therefore, it is natural to ask whether the two algebras are isomorphic.
If this is the case then $\Winf{\mu}$ must truncate to the coset orbifold~\eqref{orb} and to $\mathcal{WD}(n,n)^\tau$ 
at the same  value of $\mu$.
Let us check that this is indeed so.
First, notice that the OPE between a  $\tau$-invariant generator from the list \eqref{eq:spec_dnn} and $V$ must be linear in $V$. 
Cubic and higher powers of $V$ are not allowed to appear for dimensional reasons.
The linearity in $V$ of the above OPEs implies that $V$ generates, in the sense of sec.~\ref{sec:min}, a minimal representation of the orbifold algebra.
Comparing the conformal dimension $h=n$ of this representation with the conformal dimensions~\eqref{eq:4min_winfm_mc} of the minimal representations of $\Winf{\mu}$, we conclude that $\mu=1-2n$.
But this is the same value of $\mu$ at which $\Winf{\mu}$ truncates to $\mathcal{WD}(n,n)^\tau$, see eq.~\eqref{eq:tr_q}.
Therefore, we conclude that
\begin{equation}
\frac{\mathfrak{so}(2n+1)_k \oplus\mathfrak{so}(2n)_1}{\mathfrak{so}(2n)_{k+1}} \simeq \mathcal{WD}(n,n)\ .
\label{eq:c1_dnn}
\end{equation} 
Of course, the isomorphism holds only if we identify the central charge of the coset
\be\label{eq:cc_dnn}
c = \frac{3 k n}{2 n+k-1}
\ee 
with the central charge~\eqref{eq:expl_c_qds} of the DS reduction, i.e.\ when 
\begin{equation}\label{eq:dnn_ak}
\alpha_+^2  = 2n+k-1 \ .
\end{equation}

We can subject our conjecture~\eqref{eq:c1_dnn} to two more non-trivial checks.
Let us use the conventions of 
\cite{Creutzig:2012ar} and denote  coset representations in the Neveu-Schwarz sector for the $\mathfrak{so}(2n)_1$ fermions by $(\pi; \omega)$, 
where $\pi$ and $\omega$ are integrable weights of $\mathfrak{so}(2n+1)_k$ and $\mathfrak{so}(2n)_{k+1}$, respectively.
According to \cite{Creutzig:2012ar}, the coset representations $(f;0)$ and $(0;f)$, where $f$ is the highest weight of the vector
representation, give rise to minimal representations of $\mathcal{W}_\infty$ in the 't~Hooft limit~\eqref{eq:thooft}.
Therefore,  $(f;0)$ and $(0;f)$  are natural candidates for coset minimal representations at finite $n$ and $k$.
And indeed, we can find their  conformal dimensions
\be
h(f;0) = \frac{n}{2 n+k-1}\ ,\qquad h(0;f) = \frac{k}{2 (2 n+k-1)} \ ,
\ee
in the list of possible conformal dimensions~\eqref{eq:4min_winfm}  for the minimal representations of $\mathcal{WD}(n,n)$ 
\begin{align}\notag
h_1 &=n\ ,& h_2 &= \frac{n}{\alpha_+^2}= \frac{n}{2n+k-1}
\ ,\\
h_3 &= \frac{1-2n+\alpha_+^2}{2\alpha_+^2}=\frac{k}{2 (2 n+k-1)}
\ ,& h_4 &= \frac{1-2n+\alpha_+^2}{2} = \frac{k}{2}\ ,
\label{eq:match_conf_dim}
\end{align}
where we used eq.~\eqref{eq:dnn_ak}.
We have already seen that $h_1$ corresponds to a minimal representation of $\mathcal{WD}(n,n)^\tau$.
On the other hand,  $h_4$ does not have a natural coset interpretation.
This suggests that the minimal
representation of $\Winf{1-2n}$ of conformal dimension $h_4$ does not survive the truncation to $\mathcal{WD}(n,n)^\tau$.

At this point, one may wonder whether the other three truncations of $\mathcal{W}_\infty[\mu]$ represented in diagram~\eqref{eq:trunc_cl} also admit a  coset interpretation.
The natural guess is to look at the coset
\begin{equation}
\frac{\mathfrak{so}(2n+2)_k \oplus \mathfrak{so}(2n+1)_1}{\mathfrak{so}(2n+1)_{k+1}}
\label{c2}
\end{equation}
with central charge
\begin{equation}
c = \frac{3 k (2n+1)}{2(2 n+k)}\ .
\label{eq:cc2}
\end{equation}
The counting of generators in this case amounts to simply  replacing everywhere in sec.~\eqref{sec:counting} $2n$ with $2n+1$ and changing the expressions in~\eqref{eq:bilnn} to
\begin{equation*}
(V^0\partial^{2m+1}V^0)\ ,\quad (V^1\partial^{2m}V^1)\ ,\quad (V^0\partial^mV^1)\ , \quad m = 0,1,2,\dots
\end{equation*}
because $V^0$ and $V^1$ change statistics.
With these modifications, the spectrum  of coset generators becomes
\begin{equation}\label{spec2}
(\tfrac{3}{2},2) \  , \ (2,\tfrac{5}{2}) \  , \ (\tfrac{7}{2},4)\  , \ \dots \  , \ (2n,2n+\tfrac{1}{2}) \  , \ (n+\tfrac{1}{2},n+1) \ ,
\end{equation}
which  agrees precisely with the algebra $\mathcal{WD}(n+1,n)$. Therefore, we conjecture that 
\begin{align}\label{i2}
\frac{\mathfrak{so}(2n+2)_k \oplus \mathfrak{so}(2n+1)_1}{\mathfrak{so}(2n+1)_{k+1}} &\cong \mathcal{WD}(n+1,n)
\end{align}
when the central charges \eqref{eq:cc2} and \eqref{eq:expl_c_qds} agree, i.e.\ when
\begin{equation}
\alpha_+^2 = 2n+k\ .
\label{eq:ac2}
\end{equation}
The  isomorphism \eqref{i2} passes the same checks as~\eqref{eq:c1_dnn}, i.e.\
the list \eqref{eq:4min_winfm} of possible conformal dimensions for the minimal representations of $\mathcal{WD}(n+1,n)$, which can also be obtained from \eqref{eq:match_conf_dim}
by shifting $n\mapsto n+\tfrac{1}{2}$, contains
the conformal dimensions of the coset orbifold representation generated by the last current in~\eqref{spec2} together with the two coset representations $(f;0)$ and $(0;f)$.

There is one last natural coset to consider
\begin{align}\label{c3}
\frac{\mathfrak{osp}(1\vert 2n)_k \oplus \mathfrak{sp}(2n)_{-1}}{\mathfrak{sp}(2n)_{k-1}}\ ,
\end{align}
which (in our conventions for the level $k$) has central charge
\begin{align}\label{eq:cc3}
c &= - \frac{3 k n}{2 n+k+1}\ .
\end{align}
This coset is similar to the previous ones because in the $k\to\infty$ limit it reduces again to a  singlet algebra of free fields, but this time they transform in the vector representation of $\mathfrak{sp}(2n)$.
Instead of the Majorana  fermions $\psi^i$ we now have $2n$ \emph{bosonic} ghosts $\beta^i$ of conformal dimension $h=1/2$
\begin{equation}
\beta^i(z)\beta^j(w) = \beta^j(w)\beta^i(z) \sim \frac{\epsilon^{ij}}{z-w}\ ,
\label{eq:bg-gh}
\end{equation}
while the role of the bosonic currents $\tilde{J}^i$ is now taken by   $2n$ \emph{fermionic} ghost currents $\xi^i$ of conformal dimension $h=1$
\begin{equation}
\xi^i(z)\xi^j(w)=-\xi^j(w)\xi^i(z)\sim \frac{\epsilon^{ij}}{(z-w)^2}\ .
\label{eq:sf-gh}
\end{equation}
Here $\epsilon^{ij}$ is a non-degenerate antisymmetric matrix defining an $\mathfrak{sp}(2n)$ invariant scalar product.
We shall denote its inverse by $\epsilon_{ij}$.
Again, classical invariant theory tells us that, in the $k\to\infty$ limit, one can take as coset generators the elementary invariants
\begin{equation}
\epsilon_{ij} (\xi^i\partial^{2m}\xi^j)\ ,\quad \epsilon_{ij} (\beta^i\partial^{2m+1}\beta^j)\ ,\quad \epsilon_{ij} \xi^i\partial^{2m}\beta^j\ ,\quad m= 0,1,2\dots\ .
\label{eq:gen_bg}
\end{equation}
With respect to the $\mathcal{N}=1$ subalgebra
$$
G \propto \epsilon_{ij}\beta^i \xi ^j\ ,\quad T = \frac{1}{2}\epsilon_{ij} (\xi^i \xi^j) -  \frac{1}{2}\epsilon_{ij} (\beta^i \partial \beta^j)
$$
they organize again according to the list \eqref{eq:Winf_spec}.
Thus,  the coset algebra~\eqref{c3} must be a quotient of $\mathcal{W}_\infty$, a fact which was already pointed out in \cite{Creutzig:2012ar}.
Notice that there is no analog of the automorphism $\tau$ in this case.

Clearly, there are relations between the  infinite number of generators \eqref{eq:gen_bg}, because there is only a finite number of fields $\beta^i,\xi^i$.
Such relations start appearing at spin $s\geq 2n+\tfrac{3}{2}$ and the  first few of them, ordered according to the spin, are simply a consequence of the fermionic nature of $\xi^i$
$$
:(\epsilon_{ij}\xi^i \xi^j)^{n} \epsilon_{kl}\xi^k\beta^l\!:\ =0\ ,\quad :(\epsilon_{ij}\xi^i \xi^j)^{n+1}\!:\ =0\ ,\quad  :(\epsilon_{ij}\xi^i \xi^j)^{n-1}\epsilon_{kl}\beta^k\xi^l\epsilon_{rs}\xi^r\partial\beta^s\!:\ =0\ \; \text{etc.}
$$
Here  we have used columns to denote the standard free field normal ordering.
The systematic way to control these relations is provided by  the second fundamental theorem of classical invariant theory \cite{Weyl}, see \cite{deBoer:1993gd} for an example of such an analysis.
The upshot is that if one rewrites them in terms of the generators~\eqref{eq:gen_bg} and the normal ordering~\eqref{eq:no}, then we get an expression for the generators
of spin $s\geq 2n+\tfrac{3}{2}$ in terms of (linear combinations of products of) generators of lower spin.
This means that the coset algebra~\eqref{c3} is effectively generated by  the $\mathcal{N}=1$ multiplets
$$
(\tfrac{3}{2},2) \ , \ (2,\tfrac{5}{2})  \  , \ (\tfrac{7}{2},4) \  , \ \dots  \  , \ (2n, 2n + \tfrac{1}{2}) \ ,
$$
just like $\mathcal{WB}(n,n)$, see \eqref{eq:wb_spec}.
Therefore, we  conjecture that
\begin{align}\label{i3}
\frac{\mathfrak{osp}(1\vert 2n)_k \oplus \mathfrak{sp}(2n)_{-1}}{\mathfrak{sp}(2n)_{k-1}} &\cong \mathcal{WB}(n,n) \ ,
\end{align}
when the central charges \eqref{eq:cc3} and \eqref{eq:expl_c_qds} agree, i.e.\ when 
\begin{align}\label{eq:ac3}
\alpha_+^2  &= -(2n + k + 1)\ .
\end{align}
An independent check for the isomorphism~\eqref{i3} is the agreement between the  conformal dimensions of the coset representations $(0;f)$ and $(0;f)$ with
 the conformal dimensions $h_2$ and $h_3$, respectively
of the minimal representations of $\mathcal{WB}(n,n)$.
The latter can be obtained from~\eqref{eq:match_conf_dim} by changing $n\to -n$ and $k\to-k$.


\subsection{Holographic and level-rank dualities}



We have concluded in sec.~\ref{sec:coset-ds} that the $\tau$-orbifolds of the cosets (\ref{eq:c1_dnn}, \ref{i2}) and the coset \eqref{i3} are quotients of $\Winf{\mu}$,
where the values of $\mu$ and $\alpha^2_+$ in each case are written in diagram~\eqref{eq:tr_q} and eqs.~(\ref{eq:dnn_ak}, \ref{eq:ac2}, \ref{eq:ac3}).
We recall that, from a  $\Winf{\mu}$ point of view,   $\alpha^2_+$  is just a convenient parametrization of the central charge \eqref{eq:expl_c_qds}.
An important result of sec.~\ref{sec:winfmu} was that there are   generically  four different pairs $(\mu_i,\alpha_{i+})$ parametrizing isomorphic algebras $\Winf{\mu_i}$, see eq.~\eqref{eq:self_d}.
Therefore, there are also  four different values of $(n_i,k_i)$ parametrizing isomorphic (orbifolds of the) coset algebras (\ref{eq:c1_dnn}, \ref{i2}) and  \eqref{i3}.
Put differently, the structure constants of these coset algebras have the same form,  as a function of $n$ and $k$, for four different $(n_i,k_i)$.
We can derive the relations between these by inserting in eq.~\eqref{eq:self_d} the relevant values of $\mu$ and $\alpha_+^2$ collected in eqs.~(\ref{eq:tr_q}, \ref{eq:dnn_ak}, \ref{eq:ac2}, \ref{eq:ac3}).
For the cosets
\begin{equation}
\left(\frac{\mathfrak{so}(N+1)_k\oplus\mathfrak{so}(N)_1}{\mathfrak{so}(N)_{k+1}}\right)^\tau
\label{eq:c1c2tau}
\end{equation}
these can be written as
\begin{align*}
(N_1,k_1) &= (N,k)\ ,&  (N_3,k_3) &= \Big(\frac{k}{N+k-1}, \frac{N}{N+k-1}\Big)\ ,\\
 (N_2,k_2) &= (k,N)\ ,& (N_4,k_4) &= \Big(\frac{N}{N+k-1}, \frac{k}{N+k-1}\Big)\ .
\end{align*}
For the last coset~\eqref{c3}, one must replace $N\mapsto -2n$ and $k\mapsto-k$ in the above.

The invariance of the algebra~\eqref{eq:c1c2tau} under the exchange $(N,k)\mapsto(k,N)$ implies the level-rank duality
\begin{align}
\left(\frac{\mathfrak{so}(N+1)_k\oplus\mathfrak{so}(N)_1}{\mathfrak{so}(N)_{k+1}}\right)^\tau &\simeq \left(\frac{\mathfrak{so}(k+1)_N\oplus\mathfrak{so}(k)_1}{\mathfrak{so}(k)_{N+1}}\right)^\tau\ .
\label{eq:levvel_rank12}
\end{align}
Similarly, one has for $k$ even
\begin{align}
\frac{\mathfrak{osp}(1|2n)_k\oplus\mathfrak{sp}(2n)_{-1}}{\mathfrak{sp}(2n)_{k-1}} &\simeq \frac{\mathfrak{osp}(1|k)_{2n}\oplus\mathfrak{sp}(k)_{-1}}{\mathfrak{sp}(k)_{2n-1}}\ .
\label{eq:levvel_rank3}
\end{align}
To the best of our knowledge these level-rank dualities have not appeared in the literature before.
They are analogous to the level rank dualities discovered in \cite{Altschuler:1990th, Gaberdiel:2012, Candu:2012tr, Candu:2012ne}.

The invariance under the transformation $(N,k)\mapsto (N_3,k_3)$ can be rewritten as
\begin{equation}
\left(\frac{\mathfrak{so}(N+1)_k\oplus\mathfrak{so}(N)_1}{\mathfrak{so}(N)_{k+1}}\right)^\tau\simeq \mathcal{W}_\infty\left[ \frac{N-1}{N+k-1} \right]\Big/\mathcal{I}
\label{eq:hol1}
\end{equation}
for the cosets~\eqref{eq:c1c2tau}, and as
\begin{equation}
\frac{\mathfrak{osp}(1|2n)_k\oplus\mathfrak{sp}(2n)_1}{\mathfrak{sp}(2n)_{k-1}}\simeq \mathcal{W}_\infty\left[ \frac{2n+1}{2n+k+1} \right]\Big/\mathcal{I}
\label{eq:hol2}
\end{equation}
for the coset~\eqref{eq:levvel_rank3}, where $\mathcal{I}$ denotes  the maximal ideal of the corresponding $\mathcal{W}_\infty$-algebras.
This last set of equations gives  the required correspondence between the chiral algebra of the coset CFTs and the asymptotic symmetry algebra of
their holographic higher spin duals. In other words, eqs.~(\ref{eq:hol1}, \ref{eq:hol2}) show that the conjectured holographically dual pairs have, as they are supposed to, the same symmetry algebras.
The existence of such a correspondence at finite $n$ and $k$ is the first indication that the holographic duality continues to hold even in the quantum gravity regime.


\section{Conclusion}

In this work we have explained how to construct level by level (using the associativity constraints on OPEs) the  most general quantum $\mathcal{N}=1$ $\mathcal{W}_\infty$-algebra generated by a set of $\mathcal{N}=1$ current multiplets with spins~\eqref{eq:Winf_spec}.
Based on the analysis of the first few OPEs, we have concluded that, besides the central charge $c$, this algebra depends on a single additional continuous parameter $\mu$. The respective algebra was denoted by $\Winf{\mu}$, its wedge algebra was identified with a subalgebra $\wg{\mu}\subset \mathfrak{shs}[\mu]$, and its classical limit with the classical Drinfel'd-Sokolov reduction of $\wg{\mu}$, i.e.\ $\mathcal{W}_\infty^{cl}[\mu]$.
We have also carried out a similar analysis in the classical case, arriving at the conclusion that $\mathcal{W}_\infty^{cl}[\mu]$ is the most general classical $\mathcal{N}=1$ $\mathcal{W}_\infty$-algebra generated by the set of currents~\eqref{eq:Winf_spec}.
For $\mathcal{W}_\infty^{cl}[\mu]$ we have explicitly evaluated the first few Poisson brackets by the method of Jacobi identities and, on the way,  showed that  every classical $\mathcal{N}=1$ $\mathcal{W}_\infty$-algebra with generating spectrum \eqref{eq:Winf_spec} is the classical limit of some quantum $\mathcal{W}_\infty$-algebra.
Thus, $\Winf{\mu}$ can be viewed as the quantization of the asymptotic symmetry algebra of the AdS$_3$  higher spin supergravity theory of the $\mathcal{N}=1$ version 
\cite{Creutzig:2012ar} of the minimal model holography. Its construction  gives access to the correlation functions of the higher spin fields in the quantum theory.

The  agreement of symmetries between the bulk and boundary theories at finite $c$ requires the $\Winf{\mu}$ algebra on the bulk side of the duality to truncate because the $\mathcal{W}$-algebra of the cosets~\eqref{eq:n1_cosets} is necessarily  finitely generated.
Motivated by this, in a \emph{first step} we have studied finitely  generated truncations of $\Winf{\mu}$.
We have found four families of such truncations which can be  identified with the  Drinfel'd-Sokolov reductions of the Lie superalgebras 
$B(n-1,n)$ and $B(n,n)$, and
$\mathbb{Z}_2$-orbifolds of the Drinfel'd-Sokolov reductions of $D(n,n)$ and $D(n+1,n)$.
This result is summarized in diagram~\eqref{eq:tr_q} in the quantum case and diagram~\eqref{eq:trunc_cl} in the classical case.
It is  important to notice that only in the last three cases ---
$B(n,n)$, $D(n,n)$ and $D(n+1,n)$ --- do the values of $\mu$ at which the truncations happen not receive quantum corrections. In a \emph{second step}, we have performed a thorough analysis of the $\mathcal{W}$-algebra symmetries on the boundary side of the duality;
the $\mathcal{W}$-algebras of the  cosets~\eqref{eq:n1_cosets} have been identified with the previous three Drinfel'd-Sokolov reductions of $B(n,n)$, $D(n,n)$ and $D(n+1,n)$.
This central result is summarized in eq.~\eqref{eq:c1_dnn}, \eqref{i2} and \eqref{i3}.\footnote{No coset realization for the Drinfel'd-Sokolov reduction of $B(n-1,n)$ could be found.}

The agreement of bulk and boundary symmetries was then verified in eqs.~(\ref{eq:hol1}, \ref{eq:hol2}).
This suggests that the quantum theory on the bulk side of the $\mathcal{N}=1$ version of the minimal model holography \cite{Creutzig:2012ar}
must be a truncation of the Vasiliev theory which is, in the sense of \cite{Perlmutter:2012ds}, a $B(n,n)$, $D(n,n)$ or $D(n+1,n)$ Chern-Simons theory subject to AdS$_3$ asymptotic boundary conditions.
It is not clear whether the two $\mathcal{N}=1$ matter multiplets should enter the action of the truncated Vasiliev theory because the conformal dimensions of the coset states $(0;f)$ and $(f;0)$, which are dual to them in the 't~Hooft limit,
 diverge when analytically continued to large $c$ at finite $n$.
Therefore, one is tempted to interpret them as non-perturbative conical defect type classical geometries rather than fundamental fields in analogy with  \cite{Gaberdiel:2012}.

The analysis of \cite{Perlmutter:2012ds} showed that the correspondence between the  coset states (analytically continued to large $c$ and fixed $n$) and classical geometries
depends crucially on the realization of the coset theories as minimal models for the Drinfel'd-Sokolov reductions.
More precisely,  residual symmetries of classical geometries were established to be in a one to one correspondence with the
null vectors of the fully degenerate $\mathcal{W}$-algebra representations of the minimal models.
From this perspective, it would be very interesting to   reproduce the spectra of the $\mathcal{N}=1$ cosets (\ref{eq:c1_dnn}, \ref{i2}, \ref{i3}) and elucidate their  null vector structure by carrying out the analysis of the  fully degenerate representations of the Drinfel'd-Sokolov reductions of $B(n,n)$, $D(n,n)$ and $D(n+1,n)$ 
along the lines of \cite{Ito:1991, Ito:1990ac}.

In conclusion,  the $\mathcal{N}=1$ version of minimal model holography is a mixture between  the  $\mathcal{N}=2$ version \cite{Creutzig:2011fe, Candu:2012jq, Candu:2012tr} and the bosonic even spin version \cite{Gaberdiel:2011nt, Candu:2012ne}.
And although most of its features have been encountered previously, our analysis illustrates the power of the $\mathcal{W}_\infty$-algebra approach to discover new isomorphisms such as \eqref{eq:c1_dnn}, \eqref{i2} and \eqref{i3}.

\section*{Acknowledgements} 
We thank Matthias Gaberdiel, Kewang Jin and Maximilian Kelm for useful discussions and suggestions to improve the text, 
as well as Thomas Creutzig, Yasuaki Hikida and Peter R{\o}nne
for useful comments. This work is supported in parts by the Swiss National Science Foundation.



 \appendix

\section{\texorpdfstring{$\mathcal{N}=1$}{N=1} structure}\label{sec:n=1}


The $\mathcal{N}=1$ Virasoro algebra is generated by the energy momentum tensor $T(z)$ and a fermionic current $G(z)$ satisfying the  OPEs
\begin{align}\label{eq:vir1}
T(z)T(w)&\sim \frac{c}{2(z-w)^4}+\frac{2T(w)}{(z-w)^2}+\frac{\partial T(w)}{z-w}\ ,\\
T(z) G(w) &\sim  \frac{3G}{2(z-w)^2}+\frac{\partial G(w)}{z-w}\ , \notag \\
G(z)G(w)&\sim \frac{2c}{3(z-w)^3} +\frac{2T(w)}{z-w}\ . \notag
\end{align}
The Virasoro primary components $W^{s\,0}$ of spin $s$ and $W^{s\,1}$ of spin $s+\tfrac{1}{2}$ of the  $\mathcal{N}=1$ primary multiplet $W^{(s)}=\{W^{s\,0},W^{s\,1}\}$
satisfy the following OPE with the supercurrent $G(z)$
\begin{align*}
G(z)W^{s\,0}(w) &\sim  \frac{W^{s\,1}(w)}{z-w}\ ,\qquad G(z) W^{s\,1}(w)\sim \frac{2s\, W^{s\,0}(w)}{(z-w)^2}+\frac{\partial W^{s\,0}(w)}{z-w}\ .
\end{align*}
In terms of the modes
\begin{equation*}
G(z) = \sum_{n\in\mathbb{Z}+\frac{1}{2}}G_n z^{-n-\frac{3}{2}}\ ,\quad T(z) = \sum_{n\in \mathbb{Z}}L_n z^{-n-2}\ ,\quad W^{s\, \alpha}(z)=\sum_{n\in\mathbb{Z}+\frac{\alpha}{2}}W^{s\,\alpha}_n z^{-n- s-\frac{\alpha}{2}}
\end{equation*}
the  $\mathcal{N}=1$ Virasoro algebra OPEs can be rewritten as commutators
\begin{align}\label{eq:n1comm}
[L_m,L_n] &= (m-n)L_{m+n} +\tfrac{c}{12}m(m^2-1)\delta_{m+n,0}\ ,  \\
[L_m,G_r] &= \left(\tfrac{m}{2}-r\right) G_{m+r}\ ,\notag \\
[G_r,G_s] &=2L_{r+s}+\tfrac{c}{3}\left(r^2-\tfrac{1}{4}\right)\delta_{r+s,0} \ .\notag
\end{align}
Similarly, the definition of an $\mathcal{N}=1$ primary multiplet in terms of commutators is
\begin{align*}
[L_m,W^{s\,\alpha}_n] &= [m(s+\tfrac{\alpha}{2}-1)-n]W^{s\,\alpha}_{m+n}\ ,\\
 [G_m,W^{s\,0}_n] &= W^{s\,1}_{m+n}\ ,\\
[G_m,W^{s\,1}_n]&= [m(2s-1)-n]W^{s\,0}_{m+n}\ .
\end{align*}

\section{\texorpdfstring{$\W_\infty$}{Winfty} composite fields}\label{sec:composite}


Here we present a basis for the  composite $\mathcal{N}=1$ primary  fields of $\W_\infty$ up to spin~7.
In order to facilitate the construction of the basis elements,  it is convenient to introduce a ``mark'' $y_s$ for every
mode of $W^{(s)}$ that contributes to the vacuum character~\eqref{eq:ch_vac}
\begin{equation}\label{eq:ch_vac_mod}
\tilde{\chi}_\infty(q) = \prod_{s\in2\mathbb{N}-\frac{1}{2}}\prod_{n=0}^\infty\frac{1+y_sq^{s+n}}{1-y_s q^{s+\frac{1}{2}+n}}\times \prod_{s\in 2\mathbb{N}}\prod_{n=0}^\infty\frac{1+y_s q^{s+\frac{1}{2}+n}}{1-y_s q^{s+n}}\ .
\end{equation}
If we set the mark $y_{\frac{3}{2}}$ corresponding to the $\mathcal{N}=1$ Virasoro algebra modes to one and decompose the vacuum character~\eqref{eq:ch_vac_mod}
as in eq.~\eqref{eq:ch_vac}, then we get a refined generating function for the higher spin superprimaries
\begin{align}\notag
\tilde{P}(q)&=y_2 q^2 + y_{\frac{7}{2}}q^{\frac{7}{2}}+\big( y_2^2 + y_4\big)q^4+ 
  \big(y_2^2 + y_2 y_{\frac{7}{2}} + y_{\frac{11}{2}}\big)q^{\frac{11}{2}}+ \big( 
  y_2^2 + y_2^3 + y_2 y_{\frac{7}{2}} + y_2 y_4 + y_6\big)q^6\\
 &+ \big(y_2 y_\frac{7}{2} + y_2 y_4\big)q^{\frac{13}{2}}+ \big(y_2 y_\frac{7}{2} + y_2 y_4\big)q^7+\cdots\  .
 \label{eq:ref_P}
\end{align}
Thus, we immediately see that the first $\mathcal{N}=1$ primary composite field at spin 4 is essentially the normal ordered product $\left(W^{2\,0}\right)^2$.
However, the definition of normal ordering we use\footnote{The normal ordered product $(AB)(w)$ is defined in
eq.~\eqref{eq:no}, i.e.\ it is the regular term in the OPE $A(z)B(w)$.}
 does not guarantee that this field is primary. In order to obtain a primary field,
we have to add to it Virasoro descendants of the singular terms in the OPE $W^{2\,0}(z_1) W^{2\,0}(z_2)$.
The marks of these correction terms are $1$, $y_{2}$ or $y_{\frac{7}{2}}$.
We can continue this logic and associate a composite superprimary field to every term in~\eqref{eq:ref_P} that contains a product of $y_s$.
As their explicit expressions are quite involved, we shall write down only the ``dominant'' terms which have the same marks as the ones appearing in eq.~\eqref{eq:ref_P}
\begin{align} \label{eq:composite_q}
A^{(4, 0)} &= \tfrac{1}{c}\left(W^{2\, 0}\right)^2+\cdots\ , \\ \notag 
A^{(\frac{11}{2},1)} &= \tfrac{1}{c^2}G \left(W^{2\, 0}\right)^2+\tfrac{5 c+12}{48c^2} {W^{2\, 0}}' W^{2\, 1} - \tfrac{c+12}{12c^2} W^{2\, 0} {W^{2\, 1}}' + \cdots\ ,\\ \notag 
A^{(\frac{11}{2},2)} &= \tfrac{1}{c}W^{2\, 0} W^{\frac{7}{2}\, 0} +\cdots\ ,  \\ \notag 
A^{(6,1)} &= \tfrac{1}{c^2}G \left(W^{2\, 0} W^{2\, 1}\right)+\tfrac{8}{3c^2} T \left(W^{2\, 0}\right)^2 + \tfrac{10 c+233}{72c^2} {W^{2\, 0}}' {W^{2\, 0}}' - \tfrac{2 c+61}{36c^2} {W^{2\, 1}}' W^{2\, 1} \\ \notag
&\quad- \tfrac{2c+79}{18c^2} {W^{2\, 0}}'' W^{2\, 0} +
\cdots\ ,\\ \notag 
A^{(6,2)} &= \tfrac{1}{c}W^{2\, 1} W^{\frac{7}{2}\, 0} - \tfrac{4}{7c} W^{2\, 0} W^{\frac{7}{2}\, 1}  
+ \cdots\ , \\ \notag 
A^{(6,3)} &= \tfrac{1}{c}W^{2\, 0} W^{4\, 0}+ 
\cdots\ ,
\\ \notag 
A^{(6,4)} &= \tfrac{1}{c^2}W^{2\, 0} \left(W^{2\, 0}\right)^2 +\cdots\ , \\ \notag
A^{(\frac{13}{2},1)} &= \tfrac{1}{c}W^{2\, 1} W^{\frac{7}{2}\, 1}
+ \tfrac{7}{c}\, {W^{2\, 0}}' W^{\frac{7}{2}\, 0} - \tfrac{4}{c}\, W^{2\, 0} {W^{\frac{7}{2}\, 0}}' 
+\cdots\ , \\ \notag
A^{(\frac{13}{2},2)} &= \tfrac{1}{c}W^{2\, 0} W^{4\, 1} - \tfrac{2}{c} \, W^{2\, 1} W^{4\, 0}
+\cdots\ , \\ \notag
A^{(7,1)} &= \tfrac{1}{c^2}G \left( W^{2\, 0} W^{\frac{7}{2}\, 0} \right)
+ \tfrac{20c+129}{546c^2} \left( {W^{2\, 0}}' W^{\frac{7}{2}\, 1} + W^{2\, 1} {W^{\frac{7}{2}\, 0}}'  \right)
- \tfrac{4c+57}{78c^2} {W^{2\, 1}}' W^{\frac{7}{2}\, 0} \\ \notag
&\quad- \tfrac{10c+201}{546c^2} W^{2\, 0} {W^{\frac{7}{2}\, 1}}'
+\cdots\ , \\ \notag
A^{(7,2)} &= \tfrac{1}{c}W^{2\, 1} W^{4\, 1}
+ \tfrac{8}{c}  {W^{2\, 0}}' W^{4\, 0} -\tfrac{4}{c} W^{2\, 0} {W^{4\, 0}}' 
+\cdots  \ .
\end{align}
Abusing the usual notation, we have denoted by  $A^{(s,i)}$ the $\mathcal{N}=1$ primary component rather then the whole supermultiplet.
In order to get a superprimary, one must correct the  normal ordered terms that we have written down by Virasoro descendants of the singular terms in the OPEs defining these normal orderings.
The correction terms are  uniquely determined by the dominant part.

Fully explicit expressions for the classical composite $\mathcal{N}=1$ primary fields  can be obtained from the quantum ones
by first performing the  rescaling~\eqref{eq:cl_lim}
and then taking the classical limit
\begin{equation}\label{eq:comp_cl}
\tilde{A}^{(s,i)} =\lim_{\hbar\to 0} \hbar A^{(s,i)}\ .
\end{equation}
Notice that we have normalized the  fields $A^{(s,i)}$  in such a way that we get a non-zero result.
The suppressed terms in eqs.~\eqref{eq:composite_q}  will not contribute to the classical limit because they are  an effect of normal ordering.


\section{The higher spin superalgebra \texorpdfstring{$\mathfrak{shs}^\sigma[\mu]$}{shssigma}}
\label{shssigma}


In~\cite{Fradkin:1990qk}, Fradkin and Linetsky wrote down the commutation relations of  $\mathfrak{sl}(N|N-1)$ in the Racah basis\footnote{Actually, only a particular linear combination of $T^0_0$ and $U^0_0$ belongs to $\mathfrak{sl}(N|N-1)$, while another linear combination generates the center of $\mathfrak{gl}(N|N-1)$.
}
\begin{align}
T^j_m & \propto 
\sum_{m',m''} C^{(N-1)/2\, j\, (N-1)/2}_{m' m m''} E_{\underline{N/2-m''+1/2},\underline{N/2-m'+1/2}} \ ,&  (j&=0,1, \dots , N-1) \ , \notag \\
U^j_m &\propto 
\sum_{m',m''} C^{N/2 - 1 \, j\, N/2 - 1}_{m' m m''} E_{\overline{N/2-m''},\overline{N/2-m'}} \ ,& (j&=0,1, \dots , N-2) \ , \notag \\
\bar{\Psi}^j_m &\propto 
\sum_{m',m''} C^{N/2 - 1 \, j\, (N-1)/2}_{m' m m''} E_{\underline{N/2-m''+1/2},\overline{N/2-m'}} \ ,& (j&=\tfrac{1}{2},\tfrac{3}{2}, \dots , N-\tfrac{3}{2}) \ , \notag \\ 
\Psi^j_m &\propto
\sum_{m',m''} C^{(N-1)/2\, j\, N/2-1}_{m' m m''} E_{\overline{N/2-m''},\underline{N/2-m'+1/2}} \ ,& (j&=\tfrac{1}{2},\tfrac{3}{2}, \dots , N-\tfrac{3}{2}) \ ,
\label{shsPsi}
\end{align}
where  $\vert m \vert \leq j$, $C^{jj'j''}_{mm'm''}$ are the Clebsch-Gordan coefficients,  $E_{mn}$ is the standard basis of $\mathfrak{gl}(N|N-1)$
and
\begin{align*}
\underline{m} &= m = 1,\dots, N\\
\overline{m} & = m+N = N+1,\dots, 2N-1\ .
\end{align*}
When the generators~\eqref{shsPsi} are normalized appropriately, the structure constants become polynomials in $N$.
Hence, they can be deformed to continuous values of $\mu=N$ without spoiling the Jacobi identities if one removes the upper bound for $j$
in eqs.~\eqref{shsPsi}.
The resulting Lie superalgebra, which we will denote by $\mathfrak{shs}[\mu]$,
is infinite dimensional, infinite rank and generically simple.
Explicit formulas for the structure constants can be found in \cite{Fradkin:1990qk} (see also \cite{Candu:2012tr}).\footnote{Our generators $T^j_m$, $U^j_m$, $\bar{\Psi}^j_m$, $\Psi^j_m$ and the parameter $\mu$ must be identified with $\tilde{T}^j_m$, $\tilde{U}^j_m$, $\tilde{\bar{Q}}^j_m$, $\tilde{Q}^j_m$ and
$\nu+1$ from \cite{Fradkin:1990qk}.
}
From the way it was defined, it is clear that $\mathfrak{shs}[\mu]$ must acquire a proper maximal ideal $\chi_N$ at integer values of $\mu=N$ such that
$\mathfrak{shs}[\mu]$ truncates back to $\mathfrak{sl}(N \vert N-1)$
\begin{equation}\label{eq:tr_shs}
\mathfrak{shs}[\mu = N] / \chi_N = \mathfrak{sl}(N \vert N-1) \ .
\end{equation}
The ideal $\chi_N$ is spanned by the generators violating the upper bound for $j$ in eqs.~\eqref{shsPsi}.
%
%

Let us now define the automorphism $\sigma$ that fixes the subalgebra $\mathfrak{shs}^\sigma[\mu]\subset\mathfrak{shs}[\mu]$.
It is induced from an outer automorphism of $\mathfrak{sl}(N|N-1)$ defined by
\be \label{eq:sigmaComp}
\sigma ( \; \cdot \; ) = - g^{-1} ( \; \cdot \; )^{st} g \ ,
\ee
where $g$ is a similarity transformation 
\begin{align}
g_{\underline{i},\underline{j}} = (-1)^{i+1} \delta_{\underline{i},\underline{N-j+1}} \ , \quad
g_{\overline{i},\overline{j}} = (-1)^{i+1} \delta_{\overline{i},\overline{N-j}} \ , \quad
g_{\underline{i},\overline{j}} = 0 \ , \quad
g_{\overline{i},\underline{j}} = 0 \ .
\end{align}
Our convention for the  supertranspose is $(M^{st})_{ij} = M_{ji}(-1)^{|i|(|i|+|j|)}$, where $|i|=0$ for $i=1,\dots,N$ and $|i|=1$ for $i=N+1,\dots,2N-1$.
%
%
Using the symmetry property of Clebsch-Gordan coefficients
\be
C^{j_1\, j_2\, J}_{m_1 m_2 M} = (-1)^{j_2 + m_2} \sqrt{\frac{2 J + 1}{2 j_1 + 1}} C^{J\, j_2\, j_1}_{-M m_2 -m_1}  \ ,
\ee
it is straightforward to check that the outer automorphism $\sigma$ takes a particularly simple form in the Racah basis~\eqref{shsPsi}
\begin{align} \label{sigma}
 \sigma (T^{j}_m) &= (-1)^{j+1} T^{j}_m \ ,& \sigma (U^{j}_m) &= (-1)^{j+1} U^{j}_m\ , \\
\sigma (\Psi^{j}_m) &= (-1)^{j+\frac{1}{2}} \bar{\Psi}^{j}_m \ ,&
\sigma (\bar{\Psi}^{j}_m) &=  (-1)^{j+\frac{1}{2}} \Psi^{j}_m \ .
\end{align}
In this form, the automorphism $\sigma$ can be extended to $\mathfrak{shs}[\mu]$.
Thus, $\mathfrak{shs}^\sigma[\mu]$ is spanned by 
\begin{align}\label{eq:shss_span}
T^{j}_m \ , \ U^{j}_m \quad \text{ for } j\in 2\mathbb{N}-1 \qquad \text{ and }\qquad 
\Psi^{j}_m + (-1)^{j+\frac{1}{2}} \bar{\Psi}^{j}_m \quad \text{ for }  j \in \mathbb{N}-\tfrac{1}{2}\ . 
\end{align}

Having defined  $\mathfrak{shs}^\sigma[\mu]$, we can now look at its  truncations.
First, notice that because the ideal $\chi_N$ in eq.~\eqref{eq:tr_shs} is $\sigma$-invariant one can perform the following truncation
\begin{equation*}
\mathfrak{shs}^{\sigma}[N]/\chi_N^{\sigma} = \left( \mathfrak{shs}[N]/\chi_N \right)^{\sigma} 
= \mathfrak{sl}(N \vert N-1)^{\sigma} = \left\{ M \in \mathfrak{sl}(N \vert N-1) \mid M^{st} g + g M = 0 \right\} .
\end{equation*}
But this coincides, for $N$ odd, with the definition of the finite-dimensional Lie superalgebra 
$\mathfrak{osp}(N \vert N-1)$  and, for $N$ even, with
 $\mathfrak{osp}(N-1 \vert N)$, because $g$ is a graded-symmetric matrix.  Hence, we obtain  that
\begin{equation}\label{eq:shstruncc}
\mathfrak{shs}^{\sigma}[\mu=N]/\chi_N^{\sigma} = \begin{cases}
\mathfrak{osp}(N \vert N-1) & \text{for $N$ odd}\\
\mathfrak{osp}(N-1 \vert N) & \text{for $N$ even .}
\end{cases}
\end{equation}


Let us rescale the generators $T^{j}_m$, $U^{j}_m$, $\Psi^{j}_m$ and $\tilde{\Psi}^{j}_m$ by a factor
\be
\alpha^j_m = \sqrt{\frac{(j-m)!(j+m)!}{(2j)!}} \ ,
\ee
i.e.\ we define $T^{j}_m = \alpha_m^j t^{j}_m$, $U^{j}_m = \alpha_m^j u^{j}_m$, $\bar{\Psi}^{j}_m = \alpha_m^j \bar{\psi}^{j}_m$ and $\Psi^{j}_m = \alpha_m^j \psi^{j}_m$.\footnote{After this rescaling, the Clebsch-Gordan coefficients
in the commutation relations of \cite{Fradkin:1990qk} get replaced by the polynomials~\eqref{eq:uni_pol}, see \cite{Bowcock:1990ku}.}
Then the generators
\begin{equation}\label{eq:osp12}
	G_m = \psi^{1/2}_m + \bar{\psi}^{1/2}_m \ , \qquad
L_m = -\frac{t^1_m+u^1_m}{\sqrt{2}}\ ,
\end{equation}
generate an $\mathfrak{osp}(1|2)$ subalgebra satisfying the commutation relations~\eqref{eq:n1comm}.
With respect to the action of $\mathfrak{sl}(2)=\{L_0,L_{\pm1}\}$, the wedge modes of $t^j_m$, $u^j_m$, $\psi^j_m$, $\bar{\psi}^j_m$ transform  in a   representation of spin $j$.
These must group together
into representations of $\mathfrak{osp}(1|2)$ because the automorphism $\sigma$ acts trivially on~\eqref{eq:osp12}.
In sec.~\ref{sec:wedge} we have denoted by $\langle s,s+\frac{1}{2}\rangle$, $s\geq 1$ the irreducible finite dimensional representations of $\mathfrak{osp}(1|2)$;
$s=1$ corresponds to the trivial representation, while $s\geq 1$ to a representation that splits into two representations of $\mathfrak{sl}(2)$ of spin $j=s-1$ and $j=s-\tfrac{1}{2}$.
Assembling the $\mathfrak{sl}(2)$ representations~\eqref{eq:shss_span} into representations of $\mathfrak{osp}(1|2)$ we get a decomposition
\begin{equation}\label{eq:wedge}
\mathfrak{shs}^\sigma[\mu] \big\vert_{\mathfrak{osp}(1|2)}
\simeq \bigoplus_{s\in 2\mathbb{N}-\frac{1}{2}}\langle s,s+\tfrac{1}{2}\rangle \oplus  \bigoplus_{s\in 2\mathbb{N}}\langle s,s+\tfrac{1}{2}\rangle\ 
\end{equation}
that matches precisely eq.~\eqref{eq:wedgedef}.
The identification with the wedge modes of $\mathcal{W}_\infty$  is explicitly given by
\begin{align} \notag
W^{2\, 0}_m &= -\frac{(\mu+1)t^1_m+(\mu-2)u^1_m}{2 \sqrt{2} (2 \mu -1)}\ ,&
W^{2\, 1}_m &= -\frac{\sqrt{3} (\psi^{3/2}_m -  \bar{\psi}^{3/2}_m) }{2 (2 \mu -1)} \ ,\\ \notag
W^{\frac{7}{2} \, 0}_m &= -\frac{9 (\psi^{5/2}_m +  \bar{\psi}^{5/2}_m)}{4 \sqrt{10} (2 \mu -1)^{2}}  \ ,&
W^{\frac{7}{2} \, 1}_m &= \frac{9(t^3_m+u^3_m)}{4 \sqrt{5} (2 \mu -1)^{2}}\ ,\\ \notag
W^{4\, 0}_m &= -\frac{27 \left[ (\mu+3)t^3_m +(\mu-4)u^3_m\right]}{112\sqrt{5} (2 \mu -1)^3}\ ,&
W^{4\, 1}_m &= - \frac{27 (\psi^{7/2}_m -  \bar{\psi}^{7/2}) }{8 \sqrt{35} (2 \mu -1)^3}\ ,\\ \notag
W^{\frac{11}{2} \, 0}_m &= - \frac{27 (\psi^{9/2}_m +  \bar{\psi}^{9/2}_m)}{16 \sqrt{14} (2 \mu -1)^4 }\ ,&
W^{\frac{11}{2} \, 1}_m &= \frac{27 (t^5_m+u^5_m)}{16 \sqrt{7} (2 \mu -1)^4}\ ,\\ 
W^{6 \, 0}_m &= \frac{81 \left[ (\mu+5)t^5_m +(\mu-6)u^5_m\right]}{704 \sqrt{7} (2 \mu -1)^5}  \ ,&
W^{6 \, 1}_m &= - \frac{81 \sqrt{3} (\psi^{11/2}_m -  \bar{\psi}^{11/2}_m)}{32 \sqrt{154} (2 \mu -1)^5} \ , \label{cob}
\end{align}
for the first few spins, where the overall factors  arise from the normalization requirement~\eqref{eq:norm_str}.

In conclusion, let us mention that there is an equivalent definition of $\mathfrak{shs}[\mu]$ as a quotient
\be\label{shsdef}
\mathfrak{shs}[\mu] \oplus \mathbb{C} = \frac{U(\mathfrak{osp}(1|2))}{\langle \mathrm{Cas} 
- \frac{1}{4} \mu (\mu-1) {\bf 1} \rangle} \ ,
\ee
where the $\mathbb{C}$ term on the l.h.s.\ corresponds to the (coset representative of the) identity ${\bf 1}\in U(\mathfrak{osp}(1|2))$ and
 the $\mathfrak{osp}(1|2)$  Casimir is normalized as
\begin{equation}
\mathrm{Cas} = L_0^2 - \frac{1}{2}\left( L_1  L_{-1} + L_{-1}  L_{1}\right) +\frac{1}{4}\left(G_{\frac{1}{2}}G_{-\frac{1}{2}}- G_{-\frac{1}{2}}G_{\frac{1}{2}}\right)\ .
\label{eq:cas}
\end{equation}
An obvious consequence of the alternative definition~\eqref{shsdef} is the isomorphism
\begin{equation*}
\mathfrak{shs}^\sigma[\mu]\simeq \mathfrak{shs}^\sigma[1-\mu]\ .
\end{equation*}
It also shows that the two  $\mathfrak{osp}(1|2)$ Verma modules $V(h)$ with $L_0$ lowest weight $h=\mu/2$ and $h=(1-\mu)/2$ can be extended to representations of $\mathfrak{shs}^\sigma[\mu]$
--- the higher spin generators are identified with elements of $U(\mathfrak{osp}(1|2))$.

For generic $\mu$, these two Verma modules are irreducible representations of $U(\mathfrak{osp}(1|2))$, and hence of $\wg{\mu}$.
On the other hand, it is not hard to see that for  positive integer $N$, the Verma module $V((1-N)/2)$ becomes indecomposable and contains $V(N/2)$ as a submodule.
The latter can be identified with $\chi^\sigma_N\cdot V((1-N)/2)$. Therefore, the quotient space $V((1-N)/2)/V(N/2)$ is a representation of the quotient algebra $\wg{N}/\chi^\sigma_N$.
This representation has dimension $2N-1$, hence for $N$ odd it must be the vector representation of $\mathfrak{osp}(N|N-1)$ and for $N$ even the vector representation of  $\mathfrak{osp}(N-1|N)$, see
eq.~\eqref{eq:shstruncc}.


\section{Poisson brackets of \texorpdfstring{$\mathcal{W}_\infty^{cl}[\mu]$}{clWinf}}\label{app:one_more}


The Poisson brackets of $\mathcal{W}_\infty^{cl}[\mu]$, i.e.\ the classical DS reduction of $\wg{\mu}$, are of the form
\begin{equation*}
\{\tilde{W}^{s\,\alpha}(x),\tilde{W}^{s'\,\alpha'}(y)\} = C^{ss'}_{\alpha\alpha'}(\tilde{W}(y),\partial_x)\delta(x-y)\ ,
\end{equation*}
where $C^{ss'}_{\alpha\alpha'}$ are polynomials in $\partial_x$ with coefficients evaluated at $y$.
The first few of them read explicitly
\begin{align*}
C^{22}_{00}&=\tfrac{(\mu-2 ) (\mu+1 ) }{(2\mu-1 )^2}\left(\tfrac{c}{48}\partial^3+\tfrac{1}{2}T\partial -\tfrac{1}{4}\partial T\right)
-\left(W^{2\,0}\partial -\tfrac{1}{2}  \partial W^{2\,0}\right)\, ,\\
C^{22}_{01}&=\tfrac{(\mu-2 ) (\mu+1 ) }{(2 \mu-1 )^2}\left( \tfrac{3}{8}G\partial^2 - \tfrac{1}{4}\partial G\partial +\tfrac{1}{16}\partial^2 G+\tfrac{27}{8c}TG\right)
- \left(\tfrac{1}{2} W^{2\,1} \partial
-\tfrac{1}{5} \partial W^{2\,1}+\tfrac{27 }{5 c}G W^{2\,0}\right)+   W^{\frac{7}{2}\,0}\, ,\\
C^{22}_{11}&=\tfrac{(\mu-2 ) (\mu+1 ) }{(2 \mu-1 )^2}\left(
\tfrac{c}{48}\partial^4+\tfrac{5}{4}T\partial^2 - \tfrac{5}{4}\partial T \partial + \tfrac{27}{4c}T^2 + \tfrac{27}{16c} \partial G G+\tfrac{3}{8}\partial^2 T\right)
-\left(W^{2\,0}\partial^2 - \partial W^{2\,0}\partial -\right.\\
&\quad \left. \tfrac{27}{5c}G W^{2\,1}+\tfrac{54}{5c}TW^{2\,0}+\tfrac{3}{10}\partial^2 W^{2\,0}\right)
+W^{\frac{7}{2}\,1}\, , \\
C^{2\frac{7}{2}}_{00}&=
\tfrac{(\mu-3 ) (\mu+2 )}{(2 \mu-1 )^2}\left(-\tfrac{27}{80}W^{2\,1}\partial^2+ \tfrac{243}{100 c} GW^{2\,0}\partial +
\tfrac{27 }{200}\partial W^{2\,1}\partial  -\tfrac{81 }{200 c }G\partial W^{2\,0}-
\tfrac{27}{8 c }TW^{2\,1} -\right.\\
&\quad \left. \tfrac{27 }{25c }\partial G W^{2\,0} -\tfrac{9 }{400}\partial^2 W^{2\,1}\right)-
\left(\tfrac{4}{5} W^{\frac{7}{2}\,0}\partial - \tfrac{8}{35} \partial W^{\frac{7}{2}\,0}\right)
+W^{4\,1} \, ,\\
C^{2\frac{7}{2}}_{01}&=-\tfrac{(\mu-3 ) (\mu+2 )}{(2 \mu-1 )^2}\left( \tfrac{9}{20} W^{2\,0}\partial^3+\tfrac{567}{80 c}GW^{2\,1}\partial+
\tfrac{54}{5c }TW^{2\,0}\partial -
\tfrac{567}{400c } G\partial W^{2\,1}-
\tfrac{189 }{80c } \partial G W^{2\,1}-\right.\\
&\quad \left.
\tfrac{27 }{5 c } \partial T W^{2\,0}\right)
-
\left(\tfrac{4}{7} W^{\frac{7}{2}\,1}\partial + \tfrac{27 }{5 c} GW^{\frac{7}{2}\,0}-\tfrac{1}{7} \partial W^{\frac{7}{2}\,1}\right)
+\left(8 W^{4\,0}\partial- 2\partial W^{4\,0}\right)\, ,\\
C^{2\frac{7}{2}}_{10}&=\tfrac{(\mu-3 ) (\mu+2 ) }{(2 \mu -1)^2}\left(\tfrac{9}{20} W^{2\,0}\partial^3-\tfrac{27}{80}\partial W^{2\,0}\partial^2+\tfrac{783 }{50 c}TW^{2\,0}\partial+
\tfrac{1863}{400c } GW^{2\,1}\partial+
\tfrac{27 }{200}\partial^2W^{2\,0}\partial -\right.\\
&\quad \left.\tfrac{81}{80 c} G\partial W^{2\,1}-
\tfrac{837 }{200 c} T\partial W^{2\,0}-
\tfrac{297 }{100 c } \partial G W^{2\,1}-
\tfrac{189 }{25 c }\partial T W^{2\,0}-
\tfrac{9 }{400 } \partial^3 W^{2\,0}\right)-
\left(
\tfrac{8}{35} W^{\frac{7}{2}\,1}\partial - \right.\\
&\quad \left. \tfrac{27 }{5c} GW^{\frac{7}{2}\,0}-\tfrac{3}{35} \partial W^{\frac{7}{2}\,1}\right)-
\left(8 W^{4\,0}\partial-3\partial W^{4\,0}\right)\, ,\\
 C^{2\frac{7}{2}}_{11}&=\tfrac{(\mu-3 ) (\mu+2 )}{(2\mu-1)^2}\left(-\tfrac{63}{80} W^{2\, 1}\partial^3+ \tfrac{243 }{100 c } GW^{2\,0}\partial^2+
\tfrac{189 }{400} \partial W^{2\,1}\partial^2-
\tfrac{567 }{20 c} TW^{2\,1}\partial +
\tfrac{1701 }{400c } G \partial W^{2\,0}\partial -\right.\\
&\quad \left.
\tfrac{891}{100 c } \partial G W^{2\,0}\partial -
\tfrac{63}{400 }\partial^2 W^{2\,1}\partial
-\tfrac{81}{80 c } G\partial^2 W^{2\,0}+\tfrac{621}{100c} T\partial W^{2\,1}
-\tfrac{351 }{400c} \partial G\partial W^{2\,0}+\right.\\
&\quad \left.
\tfrac{27}{2 c} \partial T W^{2\,1}+
\tfrac{189}{50 c} \partial^2 G W^{2\,0}
+\tfrac{9 }{400}\partial^3 W^{2\,1}
\right)
-\left(
\tfrac{4}{5} W^{\frac{7}{2}\,0}\partial^2- 
\tfrac{16}{35} \partial W^{\frac{7}{2}\,0}\partial 
-\tfrac{27}{5 c}  GW^{\frac{7}{2}\,1}+\right.\\
&\quad \left.
\tfrac{54 }{5 c} TW^{\frac{7}{2}\,0}+
\tfrac{3}{35}\partial^2 W^{\frac{7}{2}\,0}\right)
+\left(
9 W^{4\,1}\partial-
3 \partial W^{4\,1}\right)\, ,
\end{align*}
where to lighten the notation we have dropped the tildes from $G,T$ and $W^{s\,\alpha}$.


\section{Simple root systems and weights}\label{sec:lsa}


In this appendix we shall write down for the Lie superalgebras
\begin{align} \notag
B(n,n) &= \mathfrak{osp}(2n+1|2n) \ , & B(n-1,n) &= \mathfrak{osp}(2n-1|2n) \ , \\ \label{eq:superb}
D(n,n) &= \mathfrak{osp}(2n|2n) \ , & D(n+1,n) &= \mathfrak{osp}(2n+2|2n) \ ,
\end{align}
the root system, the purely fermionic simple root system, the standard orthogonal weight space basis, our preferred normalization for it,
 the Weyl vector and covector, and the highest weight of the vector representation.
Most of the formulas are taken from \cite{Evans:1990qq}.
Let us start with two definitions.

The \emph{Weyl vector} and \emph{covector} of a Lie superalgebra w.r.t.\ a simple root system $\{\alpha_i\}_{i=1}^r$, where $r$ is the rank, are defined by the equations
\be
( \rho^{\vee} , \alpha_i ) = 1 \ , \qquad ( \rho , \alpha_i ) = \frac{(\alpha_i , \alpha_i)}{2}\ . 
\ee

A \emph{dominant weight} is by definition the highest weight of a finite dimensional representation.
For a weight of a Lie superalgebra to be dominant, it is necessary and sufficient that it has non-negative Dynkin labels w.r.t.\ the bosonic subalgebra of the superalgebra, i.e.
\be
\bigl( \Lambda , \beta_i \bigr) \geqslant 0 \ ,
\ee
where $\{\beta_i\}_{i=1}^r$ is the simple root system of the bosonic subalgebra of the superalgebra.
We shall now consider the Lie superalgebras~\eqref{eq:superb} case by case.

$a)$ $B(n,n) = \mathfrak{osp}(2n+1|2n)$,  bosonic subalgebra $\mathfrak{osp}(2n+1|2n)_0 = \mathfrak{so}(2n+1) \oplus \mathfrak{sp}(2n)$.\\
The bosonic and fermionic root systems are
\be
\Delta_0 = \{\pm \epsilon_i \pm \epsilon_j,\epsilon_i\}_{i, j=1,\, i\neq j}^n\cup\{ \pm \delta_i \pm \delta_j\}_{i,j=1}^n\ ,\quad \Delta_1 = \{\pm \epsilon_i \pm \delta_j, \pm\delta_j\}_{i,j=1}^{n}\ .
\ee
The simple root system is 
\begin{align*}
\{\alpha_{2i-1} = \epsilon_i - \delta_i\}_{i=1}^n\cup\{\alpha_{2i}=\delta_i - \epsilon_{i+1}\}_{i=1}^{n-1}\cup \{\alpha_{2n}=\delta_n\}\ .
\end{align*}
Scalar product in weight space
$$
(\epsilon_i,\epsilon_j)= +\delta_{ij}\ ,\qquad (\delta_i,\delta_j)=-\delta_{ij}\ .
$$
The  Weyl vector and covector are
$$
\rho =  \sum_{i=1}^n\tfrac{1}{2}(\delta_i-\epsilon_i) \ , \qquad \rho^{\vee} = \sum_{i=1}^n(2 n-2i+2) \epsilon_i - \sum_{i=1}^n(2n-2i+1) \delta_i\ .
$$
Highest weight of the vector representation $v = \epsilon_1$.

$b)$ $B(n-1,n) = \mathfrak{osp}(2n-1|2n)$, bosonic subalgebra $\mathfrak{osp}(2n-1|2n)_0 = \mathfrak{so}(2n-1) \oplus \mathfrak{sp}(2n)$.\\ The bosonic and fermionic root systems are
$$
\Delta_0 = \{\pm \epsilon_i \pm \epsilon_j,\epsilon_i\}_{i, j=1,\, i\neq j}^{n-1}\cup\{ \pm \delta_i \pm \delta_j\}_{i,j=1}^n\ ,\quad \Delta_1 = \{\pm \epsilon_i \pm \delta_j, \pm\delta_j\}_{i=1,j=1}^{n-1,n}\ .
$$
The fermionic simple root system is
\begin{align*}
\{\alpha_{2i-1} = \delta_i - \epsilon_i\}_{i=1}^{n-1}\cup\{\alpha_{2i}=\epsilon_i - \delta_{i+1}\}_{i=1}^{n-1}\cup \{\alpha_{2n-1}=\delta_n\}\ .
\end{align*}
Scalar product in weight space
$$
(\epsilon_i,\epsilon_j)= -\delta_{ij}\ ,\qquad (\delta_i,\delta_j)=+\delta_{ij}\ .
$$
The Weyl vector and covector are
$$
\rho =  \sum_{i=1}^n\tfrac{1}{2}\delta_i-\sum_{i=1}^{n-1}\tfrac{1}{2}\epsilon_i \ , \qquad \rho^{\vee} = -\sum_{i=1}^{n-1}(2 n-2i) \epsilon_i + \sum_{i=1}^n(2n-2i+1) \delta_i\ .
$$
Highest weight of the vector representation $v = \delta_1$.

$c)$ $D(n,n) = \mathfrak{osp}(2n|2n)$,  bosonic subalgebra $\mathfrak{osp}(2n|2n)_0 = \mathfrak{so}(2n) \oplus \mathfrak{sp}(2n)$.\\
The bosonic and fermionic root systems are 
$$\Delta_0 = \{\pm \epsilon_i \pm \epsilon_j\}_{i, j=1,\, i\neq j}^n\cup\{ \pm \delta_i \pm \delta_j\}_{i,j=1}^n\ ,\quad \Delta_1 = \{\pm \epsilon_i \pm \delta_j\}_{i,j=1}^n\ .$$
The fermionic simple root system is 
$$\{\alpha_{2i-1} = \delta_i - \epsilon_i\}_{i=1}^n\cup  \{\alpha_{2i} = \epsilon_i - \delta_{i+1}\}_{i=1}^{n-1}\cup \{\alpha_{2n} = \delta_n + \epsilon_n\} \ .$$
Scalar product in weight space
$$
(\epsilon_i,\epsilon_j)= -\delta_{ij}\ ,\qquad (\delta_i,\delta_j)=+\delta_{ij}\ .
$$
The  Weyl vector and covector are
$$\rho = 0 \ , \qquad \rho^{\vee} =  - \sum_{i=1}^n(2n-2i) \epsilon_i +\sum_{i=1}^n (2n-2i+1)\delta_i\ .$$
Highest weight of the vector representation $v = \delta_1$.

$d)$  $D(n+1,n) = \mathfrak{osp}(2n+2|2n)$, bosonic subalgebra $\mathfrak{osp}(2n+2|2n)_0 = \mathfrak{so}(2n+2) \oplus \mathfrak{sp}(2n)$.
The bosonic and fermionic root systems are 
$$\Delta_0 = \{\pm \epsilon_i \pm \epsilon_j\}_{i, j=1,\, i\neq j}^{n+1}\cup\{ \pm \delta_i \pm \delta_j\}_{i,j=1}^n\ ,\quad \Delta_1 = \{\pm \epsilon_i \pm \delta_j\}_{i=1,j=1}^{n+1,n}\ .$$
The fermionic simple root system is 
$$\{\alpha_{2i-1} = \epsilon_i - \delta_i\}_{i=1}^n\cup  \{\alpha_{2i} = \delta_i - \epsilon_{i+1}\}_{i=1}^{n}\cup \{\alpha_{2n+1} = \delta_n + \epsilon_{n+1}\} \ .$$
Scalar product in weight space
$$
(\epsilon_i,\epsilon_j)= +\delta_{ij}\ ,\qquad (\delta_i,\delta_j)=-\delta_{ij}\ .
$$
The Weyl vector and covector are
$$\rho = 0 \ , \qquad \rho^{\vee} =  \sum_{i=1}^{n+1}(2n-2i+2) \epsilon_i -\sum_{i=1}^n (2n-2i+1)\delta_i\ .$$
Highest weight of the vector representation $v = \epsilon_1$.

\end{document}